**Arbitrary Reflectionless Optical Routing via Non-Hermitian Zero-Index Networks**

*Yongxing Wang*, Zehui Du, Zhenshuo Xu, Pei Xiao, Jizi Lin, Yufeng Zhang, and Jie Luo**

Y. Wang, Z. Du, Z. Xu, P. Xiao, J. Lin,

Zhangjiagang Campus, Jiangsu University of Science and Technology, Zhangjiagang 215600, China

Email: 201900000107@just.edu.cn

Y. Wang, P. Xiao, J. Lin, Y. Zhang

Department of Physics, Jiangsu University of Science and Technology Suzhou Institute of Technology, Zhangjiagang 215600, China

J. Luo

School of Physical Science and Technology, Soochow University, Suzhou 215006, China

Email: luojie@suda.edu.cn

J. Luo

Jiangsu Physical Science Research Center, Nanjing 210093, China

Optical routers are fundamental to photonic systems, but their performance is often limited by unwanted reflections and constrained functionalities. Existing design strategies generally lack complete control over reflectionless pathways and typically require computationally intensive iterative optimization. A general analytical framework for the inverse design of arbitrary reflectionless routing has remained unavailable. Here, we present an analytical inverse-design approach based on non-Hermitian zero-index networks, which enables arbitrary reflectionless routing for nearly any reciprocal scattering response. By establishing a direct algebraic mapping between target scattering responses and the network's physical parameters, we transform the design process from iterative optimization into deterministic calculation. This approach enables the precise engineering of arbitrary reflectionless optical routing. We demonstrate its broad utility by designing devices from unicast and multicast routers with full amplitude and phase control to coherent beam combiners and spatial mode demultiplexers in four-port and six-port networks. Our work provides a systematic and analytical route to designing advanced light-control devices.

## 1. Introduction

Optical routers are fundamental components for directing light in advanced photonic circuits. Platforms such as integrated silicon photonics [1–4], photonic processors [5,6], topological structures [7–9], and metasurfaces [10–12] have been extensively developed for optical routing. A fundamental and persistent challenge across these systems, however, is the suppression of unwanted reflections, which degrade device performance by introducing crosstalk and signal interference. Recently, the theory of reflectionless scattering modes (RSMs), which generalizes earlier concepts including coherent perfect absorption (CPA) and critical coupling, has provided a powerful framework to achieve reflectionless routing [13–19]. Further strategies have been proposed to mitigate the wavefront sensitivity of these approaches, i.e., the need for precise amplitude and phase combinations across the input ports, by exploiting RSM degeneracies [19], designing wavefront-robust CPA [20], or incorporating external anti-reflection structures [21]. Despite these advances, creating devices that are simultaneously reflectionless and capable of arbitrary routing remains challenging. Moreover, current inverse-design approaches typically rely on iterative numerical optimization and lack general analytical guidance for achieving arbitrary target scattering responses [15,22–28].

In this work, we propose an analytical inverse-design approach based on a customizable $N$-port non-Hermitian zero-index material (ZIM) network that enables arbitrary reflectionless routing for almost any desired reciprocal scattering response. ZIMs, characterized by near-zero permittivity and/or permeability, have been demonstrated to exhibit unique wave phenomena [29–33] and support rich non-Hermitian physics, including CPA [34–36], exotic transmission and scattering [37–39], and exceptional points [40–42]. The present ZIM network allows us to establish a direct algebraic mapping between the target scattering matrix and the network's characteristic parameters. This mapping forms the foundation of our analytical inverse-design approach, transforming the design task from computationally intensive iterative optimization into deterministic and direct calculation for any desired symmetric scattering matrix.

Crucially, this approach enables the inverse design of networks characterized by nearly arbitrary symmetric scattering matrices, including those with zero submatrices, thereby achieving wavefront-robust routing and overcoming the input-sensitivity limitations of conventional RSMs [13,14]. We demonstrate the power and generality of our analytical inverse-design approach through the systematic design of multiple functional devices in four-port and six-port ZIM networks, including unicast and multicast routers with full amplitude and phase control, coherent beam combiners, and spatial mode demultiplexers. Our work establishes a direct analytical pathway for engineering complex, functional, and robust wave-control devices, paving the way toward advanced photonic circuits.

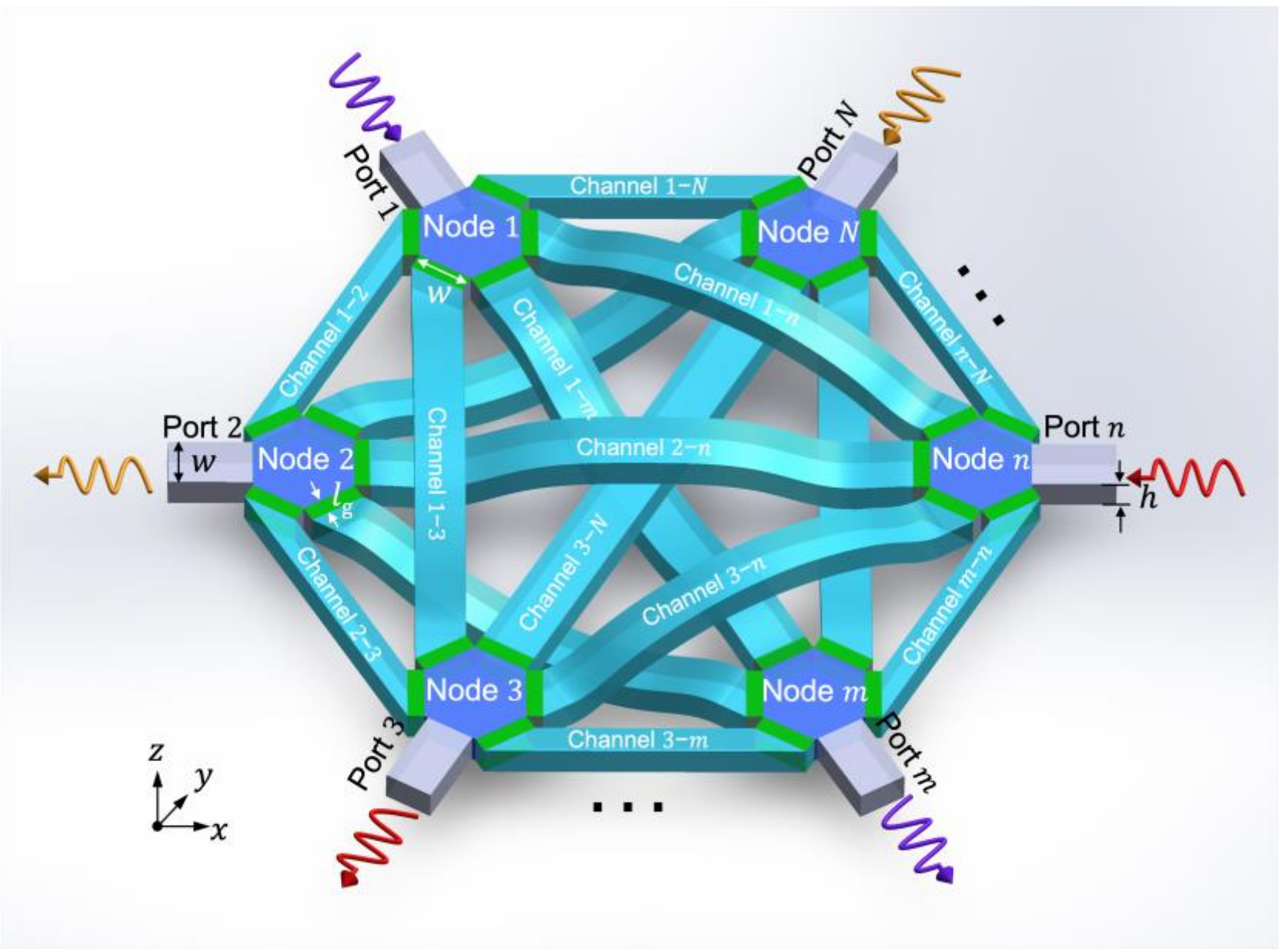

**Figure 1.** Schematic of a customizable $N$-port reflectionless optical router based on a non-Hermitian ZIM network. The network comprises $N$ engineered ZIM nodes (each connected to an input/output port) coupled pairwise via ZIM channels. Engineered dielectric gaps (green regions) separate the nodes from the channels.

## 2. Theory and model of arbitrary reflectionless optical routing

The operation of an $N$-port optical router is governed by its $N \times N$ scattering matrix $\boldsymbol{S}$, which linearly relates the complex input wave amplitudes $\boldsymbol{a}$ and output amplitudes $\boldsymbol{b}$ via $\boldsymbol{b} = \boldsymbol{S}\boldsymbol{a}$ [43,44]. The matrix element $S_{mn}$ defines the transmission coefficient from input port $n$ to output port $m$. Thus, the design of an optical router corresponds to engineering a desired scattering matrix $\boldsymbol{S}$.

To achieve arbitrary reflectionless routing, we propose an $N$-port non-Hermitian ZIM network, as illustrated in Fig. 1. This network consists of $N$ engineered ZIM nodes and $N(N-1)/2$ ZIM channels (width $w$, height $h$). Each node connects an input/output port (relative permittivity $\varepsilon_p$), and every node pair is coupled via a ZIM channel. The channels are spatially staggered and do not directly intersect, forming an overpass-like structure. All ZIM components (both nodes and channels) possess a near-zero relative permittivity ($\varepsilon \approx 0$) and a complex relative permeability $\mu_{mn}$, where $\mu_{mn}$ corresponds to a node when $m = n$ and to the channel linking the $m$-th and $n$-th ports when $m \neq n$. A key design feature is the inclusion of dielectric gaps (relative permittivity $\varepsilon_g$, length $l_g$) placed between nodes and channels. These gaps break the global field uniformity inherent to ZIMs, thereby unlocking the network's full design freedom [41]. The entire network is confined within a waveguide with perfect electric conductor (PEC) sidewalls and perfect magnetic conductor (PMC) top and bottom walls, supporting a transverse electromagnetic mode with its magnetic field oriented along the $z$ direction. Although ideal PMC boundaries are assumed here for the ideal ZIM model, they can be approximated experimentally using approaches such as high-impedance surfaces [45-47], grooved metallic structures [48], all-dielectric metasurfaces supporting magnetic resonances [49-51], or PEC walls placed a quarter wavelength away from the ZIM [34, 52].

For this ZIM network, we derive a compact algebraic relation that directly links $\boldsymbol{S}$ to the network's physical parameters through an auxiliary matrix $\boldsymbol{W}$ (see Supporting Information, section 1):

$$\boldsymbol{W} = -2i(\boldsymbol{S} + \boldsymbol{I})^{-1}. \quad (1)$$

The elements $W_{mn}$ are functions of a set of dimensionless, complex valued characteristic parameters $\xi_{mn}$, defined by $\xi_{mn} \equiv k_0 \mu_{mn} A_{mn} / z_p w$, which encapsulate the geometric and electromagnetic parameters of each component. Here, $k_0$ is the free-space wavenumber; $A_{mn}$ is the area of the corresponding network component's top or bottom surface (nodes for $m = n$, channels for $m \neq n$). $z_p = \sqrt{1/\varepsilon_p}$ is the relative impedance of ports.

Equation (1) serves as the foundation of our analytical inverse-design approach. Solving it yields explicit formulas for each $\xi_{mn}$ in terms of the target scattering matrix $\boldsymbol{S}$:

$$\xi_{nn} = \frac{-2i\sum_{i=1}^{N}(1-2\delta_{in})C_{ni}(\boldsymbol{S}+\boldsymbol{I})}{\det(\boldsymbol{S}+\boldsymbol{I})} + (N-1)\cot(k_g l_g) - i, \quad (2a)$$

$$\xi_{mn} = \frac{i\det(\boldsymbol{S}+\boldsymbol{I})}{2\sin^2(k_g l_g)\, C_{mn}(\boldsymbol{S}+\boldsymbol{I})} + 2\cot(k_g l_g), (m \neq n), \quad (2b)$$

where $\det(\boldsymbol{S}+\boldsymbol{I})$ and $C_{mn}(\boldsymbol{S}+\boldsymbol{I})$ are the determinant and the ($m$,$n$)-cofactor of the matrix $\boldsymbol{S}+\boldsymbol{I}$, respectively. $k_g = \sqrt{\varepsilon_g} k_0$ is wavenumber in gaps. For any desired reflectionless routing function described by a specific scattering matrix $\boldsymbol{S}$, provided $\boldsymbol{S}+\boldsymbol{I}$ is non-singular, Eqs. (2a) and (2b) allow direct calculation of all parameters $\xi_{mn}$. From these, all necessary geometric and electromagnetic parameters ($\mu_{mn}$, $A_{mn}$, and $w$) can be determined directly via $\xi_{mn} \equiv k_0 \mu_{mn} A_{mn} / z_p w$, thereby eliminating the need for iterative numerical optimization. We note that the ZIM network provides $N(N+1)/2$ independently tunable parameters, which exactly matches the degrees of freedom in a reciprocal $N$-port scattering matrix $\boldsymbol{S}$. This one-to-one correspondence ensures that the network configuration is uniquely determined for any given target $\boldsymbol{S}$.

It is noteworthy that our analytical approach enables the realization of nearly arbitrary symmetric scattering matrices, including those containing zero submatrices. As we elaborate below, such matrices define a reflectionless subspace that supports wavefront-robust routing, overcoming the stringent input amplitude and phase requirements that constrain conventional RSMs [13–18].

## 3. Robust reflectionless unicast routing

We begin with a four-port non-Hermitian ZIM network [Fig. 2(a)], adopting the representative parameters $\varepsilon_p = \varepsilon_g = 1$, $l_g = 0.25\lambda_0$, and $w = h = 0.5\lambda_0$, where $\lambda_0$ is the operating free-space wavelength. Our goal is to design a dual-channel, wavefront-robust, reflectionless unicast router, which supports two independent and bidirectionally reflectionless pathways with fully controllable output amplitude and phase. For routing between port pairs 1-2 and 3-4, the target scattering matrix is

$$\boldsymbol{S} = \begin{pmatrix} 0 & \alpha & 0 & 0 \\ \alpha & 0 & 0 & 0 \\ 0 & 0 & 0 & \beta \\ 0 & 0 & \beta & 0 \end{pmatrix}, \tag{3}$$

where $\alpha$ and $\beta$ are arbitrary complex transmission coefficients. Here, $S_{mn} = 0$ $(m, n = 1,3)$, indicating that ports 1 and 3 are reflectionless and mutually isolated. Applying our inverse design formalism [Eq. (2)] yields analytical expressions for the characteristic parameters:

$$\xi_{11} = \xi_{22} = i\frac{1+\alpha}{1-\alpha},\ \xi_{33} = \xi_{44} = i\frac{1+\beta}{1-\beta}, \xi_{12} = i\frac{\alpha^2-1}{2\alpha},\ \xi_{34} = i\frac{\beta^2-1}{2\beta},$$

$$\text{and}\ \xi_{13} = \xi_{14} = \xi_{23} = \xi_{24} \to \infty. \tag{4}$$

These $\xi_{mn}$ values encode the required geometry and permeability $\mu_{mn}$ of each ZIM node and channel in this network. For a fixed geometry, we have $\xi_{mn} \propto \mu_{mn}$. Under the assumed time variation term of $\mathrm{e}^{-i\omega t}$ ($\omega$ is the angular frequency), a positive imaginary part of $\xi_{mn}$ (or $\mu_{mn}$) corresponds to loss, whereas a negative one indicates gain. The extreme condition $\xi_{mn} \to \infty$ (or $\mu_{mn} \to \infty$) physically represents a PMC condition required for channel isolation. Although the required $\mu_{mn}$ values may appear extreme, they can be realized via photonic doping [33–41,53-56], as will be discussed later in the context of practical implementation. In practice, however, it is not necessary to strictly satisfy the ideal condition $\xi_{mn} \to \infty$. We find that a sufficiently large characteristic parameter $\xi_{mn}$ (typically $\xi_{mn} > 20$) is already adequate for efficient isolation. Figure 2(b) maps the accessible solution space when all ZIM components are passive or lossless ($\mathrm{Im}(\mu_{mn}) \geq 0$), showing as blue regions, plotted as a function of the amplitude and phase of $\alpha$ for different ratios $\beta/\alpha$. Such passive configurations are particularly attractive for practical applications.

As a representative example to demonstrate the remarkable ability of the ZIM network in achieving arbitrary control over both amplitude and phase, we select $\alpha = i$ and $\beta = 2$, corresponding to the point marked by a star in Fig. 2(b). Under this condition, the unicast router simultaneously achieves a $\pi/2$ phase shift in one channel and a twofold amplitude amplification in the other. Substituting these values into Eq. (4) and applying the relation $\xi_{mn} \equiv k_0 \mu_{mn} A_{mn} / z_p w$ yields the required $\mu_{mn}$ and $A_{mn}$ for each ZIM component (see

Supporting Information, sections 2.1 and 2.2). Figure 2(c) presents the simulated distribution of the normalized magnetic field $H_z/H_0$ obtained using the finite-element software COMSOL Multiphysics, when a signal with magnetic-field amplitude $H_0$ is incident at port 1. It is seen that the signal is guided to port 2 with a $\pi/2$ phase shift. Similarly, Fig. 2(d) presents the simulated $H_z/H_0$ distribution when the signal is incident at port 3, showing that the signal is routed to port 4 with doubled amplitude.

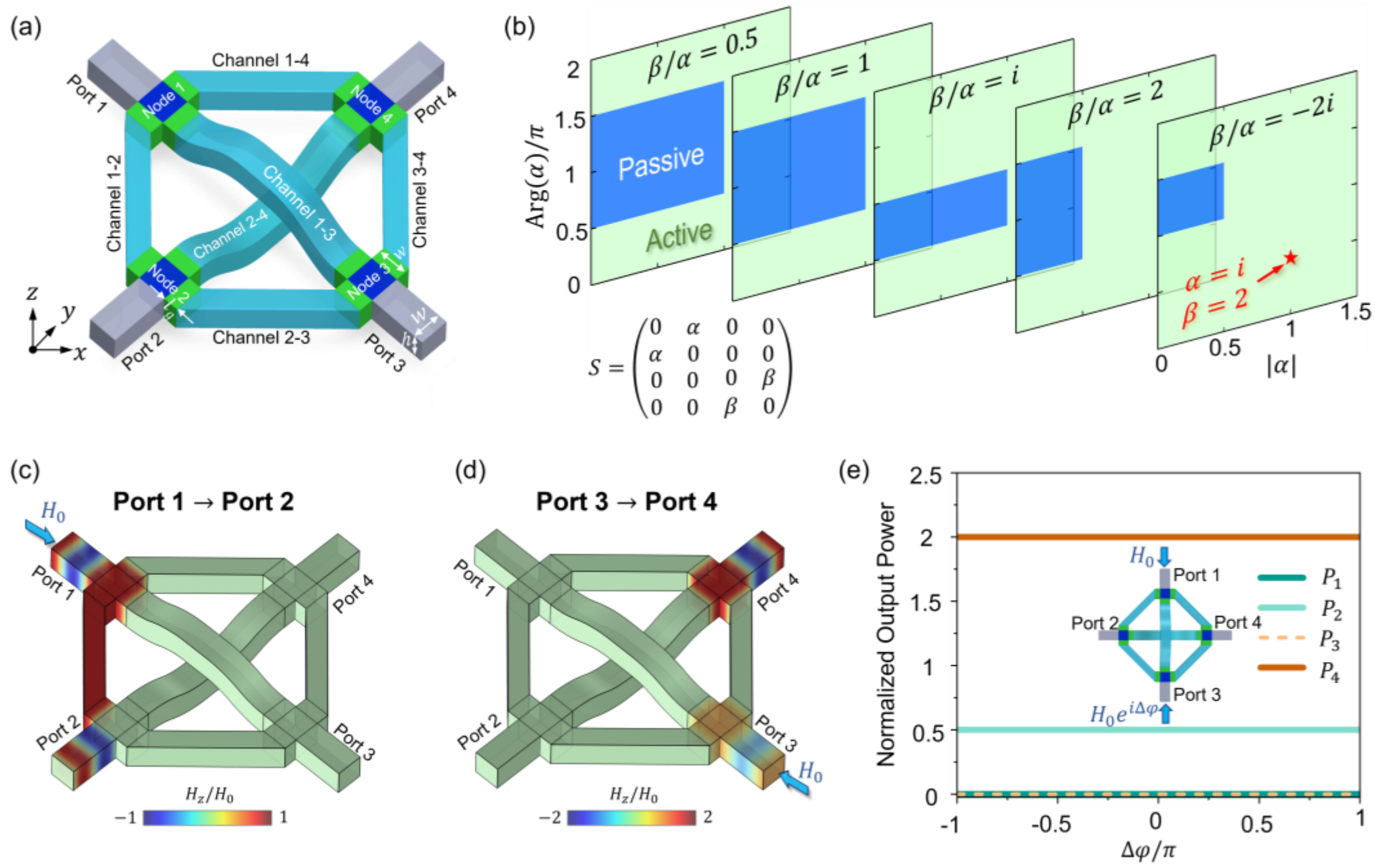

**Figure 2.** A dual-channel reflectionless unicast router. (a) Schematic of a four-port router implemented with a non-Hermitian ZIM network. (b) Design space for realizing the target scattering matrix $\boldsymbol{S}$ (inset). Blue regions indicate configurations where all ZIM components are passive or lossless ($\mathrm{Im}(\mu_{mn}) \geq 0$), while green regions require active elements ($\mathrm{Im}(\mu_{mn}) < 0$) in the parameter space of the amplitude and phase of $\alpha$ for different ratios $\beta/\alpha$. [(c) and (d)] Simulated distributions of normalized magnetic field $H_z/H_0$ for the unicast router operating at the target point ($\alpha = i$, $\beta = 2$), marked by a star in (b), when a signal is incident at (c) port 1 or (d) port 3. (e) Simulated output power $P_m$ ($m$ =1,2,3,4) at each port as a function of the relative phase difference $\Delta\varphi$ between signals simultaneously incident at ports 1 and 3. The inset illustrates the configuration.

In both cases, the input ports remain reflectionless, and the two routing channels (port 1→2,

port 3→4) are isolated from each other, achieving wavefront-robust reflectionless unicast routing. To verify this robustness, we simultaneously excite ports 1 and 3 with equal amplitude $H_0$ but a variable relative phase difference $\Delta\varphi$ [inset of Fig. 2(e)]. The simulated output power $P_m$ ($m$ =1,2,3,4) at the $m$-th port, defined as $P_m = \frac{\sum_{n=1}^{N}|S_{mn}a_n|^2}{\sum_{n=1}^{N}|a_n|^2}$ with $a_n$ being the input amplitude at the $n$-th port, is plotted as a function of $\Delta\varphi$ in Fig. 2(e). The results show $P_1 = P_2 = 0$, confirming reflectionless operation, while $P_3$ and $P_4$ are $\Delta\varphi$-independent, demonstrating both the channel isolation and wavefront-insensitive performance of the proposed non-Hermitian ZIM network.

Next, we demonstrate a practical platform for realizing such reflectionless routers. The key challenge is implementing the various ZIM components with $\varepsilon \approx 0$ and required complex $\mu_{mn}$. Our approach proceeds in two steps. First, we employ a rectangular PEC waveguide operating in the $TE_{10}$ mode near its cutoff frequency, which behaves as an effective ZIM with $\varepsilon \approx 0$ [57,58] (see Supporting Information, section 3.1). Such cutoff-waveguide implementations can operate at low frequencies, e.g., in the microwave and terahertz regimes. Second, through doping this effective ZIM with suitable dopants, we can tailor $\mu_{mn}$ across the complex plane while preserving the condition $\varepsilon \approx 0$ [33,38,41] (see Supporting Information, section 3.1).

Figure 3(a) shows the schematic of a four-port waveguide network for the dual-channel router studied in Fig. 2 with $\alpha = i$ and $\beta = 2$. The network is constructed from interconnected rectangular PEC waveguides. Air-filled waveguide sections (height $h = 0.5\lambda_0$) are operated near the $TE_{10}$-mode cutoff frequency $\omega_c$, creating an effective ZIM background with $\varepsilon \approx 0$. Within this background, the required ZIM nodes and channels are realized by introducing cylindrical loss/gain dopants (relative permittivity $\varepsilon_{d,mn}$, cross-sectional radius $R_{d,mn}$), which yield the desired effective permeability $\mu_{mn}$. The gain dopants can be implemented using established active dielectric technologies, such as dye-doped media, semiconductor quantum wells or quantum dots in the optical and infrared regimes[59], or negative-resistance-integrated metamaterials in the microwave regime [60]. We note that the waveguides linking ports 1-3 and 2-4 are spatially staggered and do not intersect directly, analogous to an overpass structure. The effect of this spatial curvature is elaborated in section 3.1 of Supporting Information. The input/output ports and coupling gaps correspond to the Teflon-filled ($\varepsilon_f = 2.1$) waveguide sections, ensuring operation above cutoff. To prevent higher-order modes, each dopant is encircled by thin PEC wires connecting the upper and bottom waveguide walls [57,58]. Based on Eq. (4) and photonic doping theory, we obtain all geometric and electromagnetic parameters ($\varepsilon_{d,mn}$, $R_{d,mn}$, $l_g$, and $w$), which are provided in sections 3.2 and 3.3 of Supporting Information.

Figure 3(b) presents the simulated S-parameter spectra, showing that at $\omega_c$, $|S_{12}| = |S_{21}| = 1$ and $|S_{34}| = |S_{43}| = 2$, while all other S-parameters are nearly zero, in full agreement with the target scattering matrix in Eq. (3) for $\alpha = i$ and $\beta = 2$. Further validation is provided by the simulated field distributions in Figs. 3(c) and 3(d), which display the normalized magnetic field $H_z/H_0$ at $\omega_c$ when port 1 or port 3 is excited, respectively. The results closely match with the ideal simulations in Figs. 2(c) and 2(e), confirming reflectionless routing from port 1 to port 2 with a $\pi/2$ phase shift and from port 3 to port 4 with twofold amplitude amplification. A quantitative analysis of the deviation between the achieved and target S-parameters is provided in section 3.7 of Supporting Information. These results demonstrate the practical feasibility of the proposed dual-channel reflectionless unicast router.

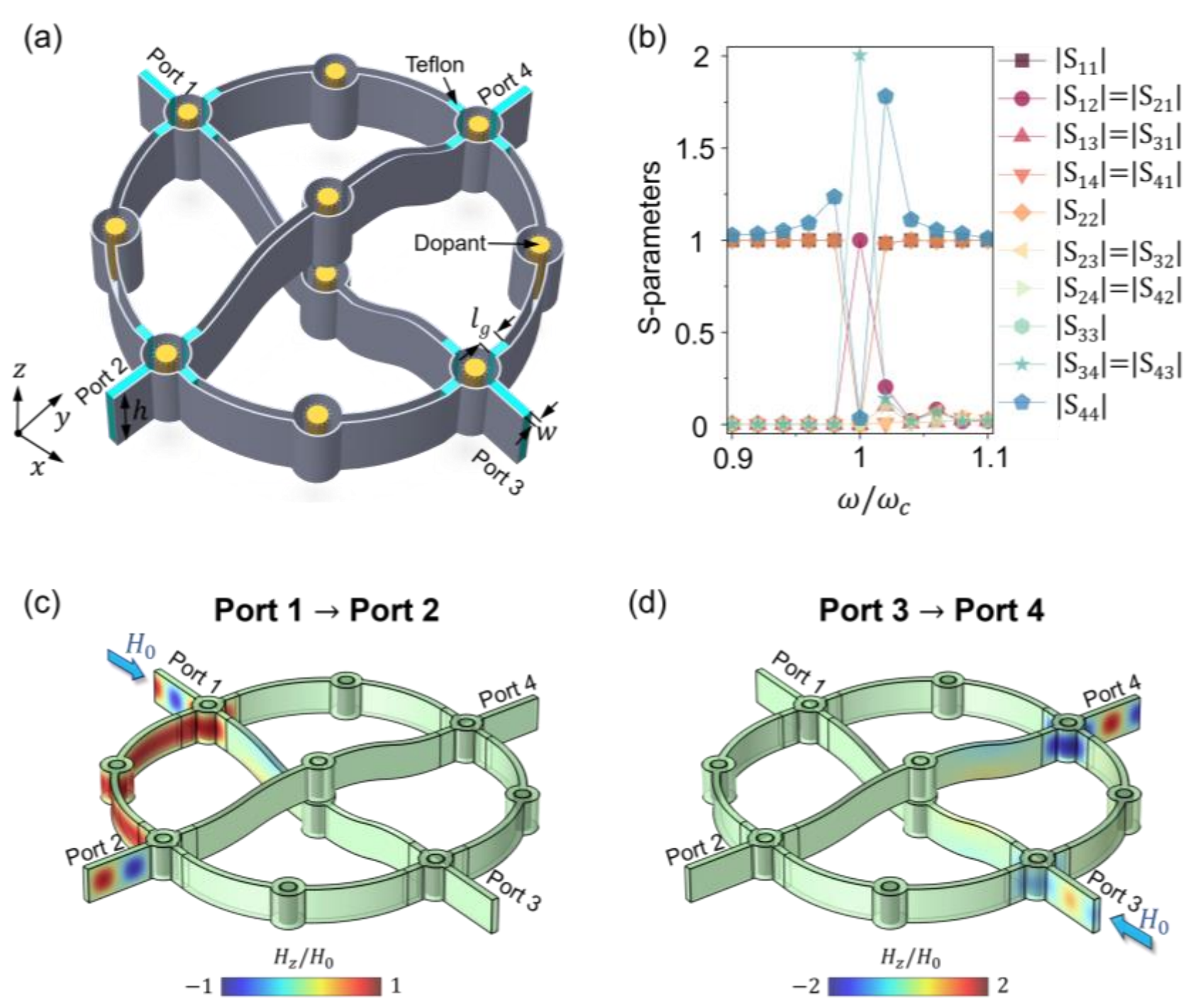

**Figure 3.** Practical implementation of a ZIM-network-based reflectionless router. (a) Schematic

of waveguide-based implementation for the unicast router studied in Fig. 2 with $\alpha = i$ and $\beta = 2$. The ZIM nodes and channels are implemented using the air-filled waveguide regions doped with tailored cylindrical inclusions, while the input/output ports and coupling gaps are formed by the Teflon-filled waveguide sections. (b) Simulated S-parameter spectra of the waveguide network, designed to realize the scattering matrix given in Eq. (3) for $\alpha = i$ and $\beta = 2$ at the operating frequency $\omega_c$. [(c) and (d)] Simulated distributions of normalized magnetic field $H_z/H_0$ at $\omega_c$ when (c) port 1 or (d) port 3 is excited.

**4. Reflectionless multicast routing and coherent beam combining**

Beyond the unicast routing demonstrated above, our non-Hermitian ZIM network also enables the analytical inverse design of reflectionless multicast routers. As a representative example, we consider routing from a single input port to two output ports within the four-port ZIM network. The corresponding target scattering matrix is,

$$\boldsymbol{S} = \begin{pmatrix} 0 & \alpha' & \beta' & 0 \\ \alpha' & 0 & 0 & 0 \\ \beta' & 0 & 0 & 0 \\ 0 & 0 & 0 & 0 \end{pmatrix}, \tag{5}$$

where $\alpha'$ and $\beta'$ are arbitrary transmission coefficients for pathways $1 \rightarrow 2$ and $1 \rightarrow 3$, respectively. Here, all other S-parameters are set to be zero, ensuring port 1 is reflectionless, ports 2 and 3 are reflectionless and mutually isolated, and port 4 remains completely decoupled. Applying our inverse design formalism [Eq. (2)] yields analytical expressions for the characteristic parameters:

$$\xi_{11} = -i\frac{\alpha'^2+2\alpha'+(1+\beta')^2}{\alpha'^2+\beta'^2-1},\ \xi_{22} = -i\frac{\alpha'^2-\beta'^2-2\alpha'(\beta'-1)+1}{\alpha'^2+\beta'^2-1},\ \xi_{33} = i\frac{\alpha'^2+2\alpha'\beta'-(1+\beta')^2}{\alpha'^2+\beta'^2-1},$$
$$\xi_{44} = i,\ \xi_{12} = i\frac{\alpha'^2+\beta'^2-1}{2\alpha'},\ \xi_{13} = i\frac{\alpha'^2+\beta'^2-1}{2\beta'},\ \xi_{23} = -i\frac{\alpha'^2+\beta'^2-1}{2\alpha'\beta'},$$
$$\text{and}\ \xi_{14} = \xi_{24} = \xi_{34} \rightarrow \infty. \tag{6}$$

The infinite values for $\xi_{14}$, $\xi_{24}$, and $\xi_{34}$ physically enforce complete isolation of port 4. From these $\xi_{mn}$ values, the required geometry and permeability $\mu_{mn}$ of each ZIM node and channel in this network can be directly determined (see Supporting Information, sections 2.1 and 2.3). Figure 4(a) maps the accessible solution space under the constraint that all ZIM components remain passive or lossless ($\mathrm{Im}(\mu_{mn}) \geq 0$), shown as blue regions, plotted as a function of the amplitude and phase of $\alpha'$ for different ratios $\beta'/\alpha'$. Such passive configurations are particularly attractive for practical implementations.

To demonstrate a general multicast case, we choose $\alpha' = -i\sqrt{1/3}$ and $\beta' = -i\sqrt{2/3}$, [marked by a star in Fig. 4(a)]. These parameters indicate that an input signal at port 1 would be split into a power ratio of $1:2$ between output ports 2 and 3, while remaining reflectionless and leaving port 4 isolated. The corresponding simulated magnetic-field distribution $H_z/H_0$ for excitation at port 1 [Fig. 4(b)] confirms the desired reflectionless multicast routing with the prescribed power division.

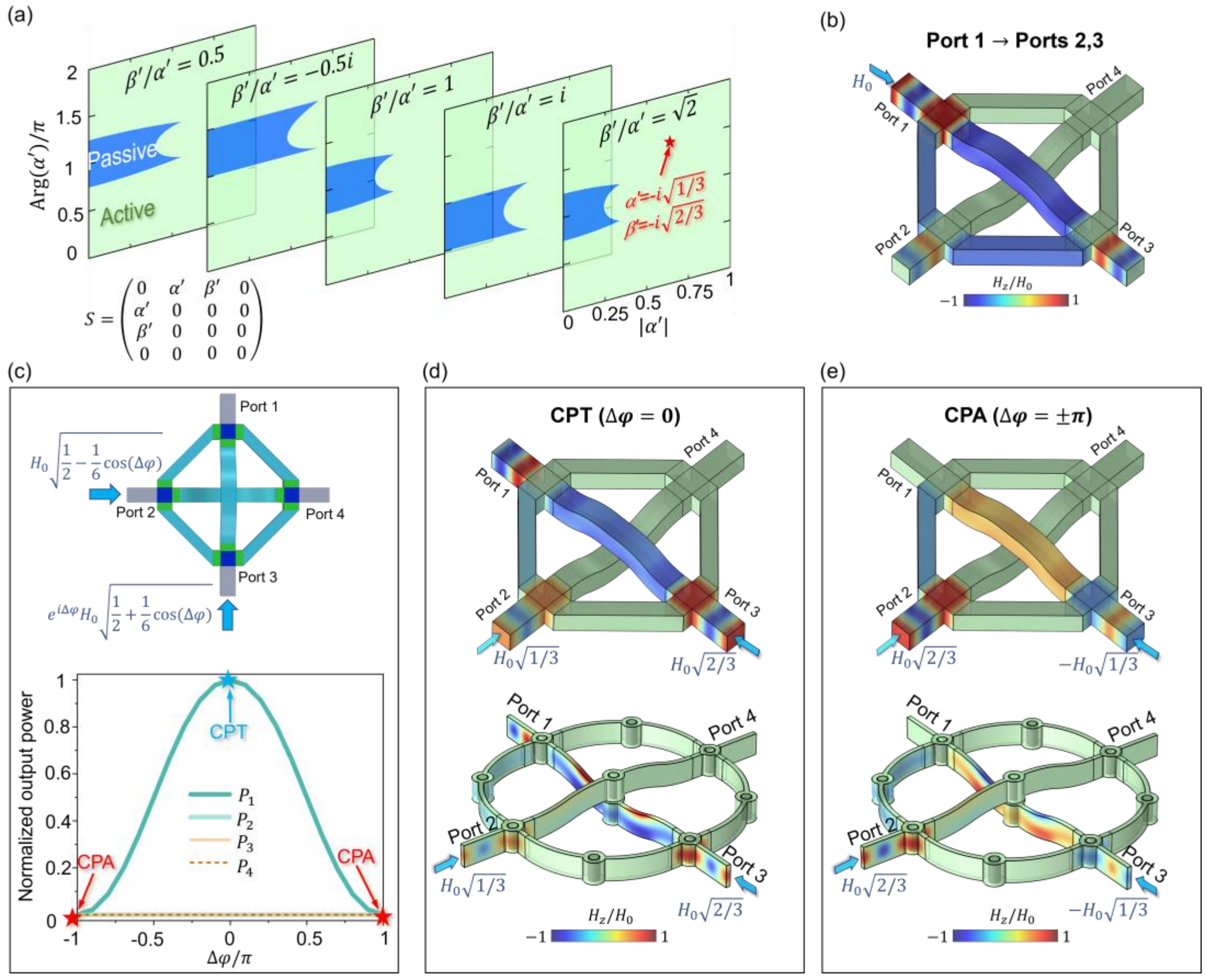

**Figure 4.** Reflectionless multicast routing and coherent signal combining. (a) Design space for realizing the target scattering matrix $\boldsymbol{S}$ (inset). Blue regions indicate configurations where all ZIM components are passive or lossless ($\mathrm{Im}(\mu_{mn}) \geq 0$), while green regions require active elements ($\mathrm{Im}(\mu_{mn}) < 0$) in the parameter space of the amplitude and phase of $\alpha'$ for different ratios $\beta'/\alpha'$. (b) Simulated magnetic-field distribution $H_z/H_0$ for the multicast router at the target point ($\alpha' = -i\sqrt{1/3}, \beta' = -i\sqrt{2/3}$), marked by a star in (a), when port 1 is excited. (c)

Upper: schematic of the reversed operation for coherent beam combining, with ports 2 and 3 simultaneously excited by tailored coherent signals. Lower: simulated output power $P_m$ ($m$ =1,2,3,4) as a function of the relative phase difference $\Delta\varphi$. [(d) and (e)] Simulated distributions of $H_z/H_0$ for the ideal ZIM network (upper) and its waveguide-based implementation operating at $\omega_c$ (lower) under (d) CPT ($\Delta\varphi = 0$) and (e) CPA ($\Delta\varphi = \pm\pi$) conditions.

Interestingly, when operated in reverse, the same network functions as a coherent beam combiner, guiding signals from ports 2 and 3 into port 1 without reflection, as indicated by $S_{22} = S_{23} = S_{32} = S_{33} = 0$ in the target scattering matrix [Eq. (5)]. The upper panel of Fig. 4(c) illustrates the configuration: tailored coherent inputs are injected into ports 2 and 3. Specifically, the incident magnetic fields at ports 2 and 3 are $a_2 = H_0\sqrt{1/2 - \cos(\Delta\varphi)/6}$ and $a_3 = H_0 e^{i\Delta\varphi}\sqrt{1/2 + \cos(\Delta\varphi)/6}$, respectively. The lower panel of Fig. 4(c) shows the simulated normalized output power $P_m$ ($m$ =1,2,3,4) as a function of relative phase difference $\Delta\varphi$. Here, $P_2 = P_3 = P_4 = 0$, independent of $\Delta\varphi$, confirming that reflectionless operation and port 4 isolation are simultaneously obtained. Notably, $P_1$ varies from unity at $\Delta\varphi = 0$ to zero at $\Delta\varphi = \pm\pi$. This behavior demonstrates coherent perfect transmission (CPT) at zero phase difference and CPA at $\pm\pi$ phase difference, achieved simply by modulating the coherent inputs.

Figure 4(d) confirms the CPT ($\Delta\varphi = 0$), comparing the distributions of $H_z/H_0$ for the ideal ZIM network (upper) with its waveguide-based implementation operating at $\omega_c$ (lower). Both show that nearly all incident power is guided reflectionlessly into port 1. Conversely, Fig. 4(e) shows the CPA case ($\Delta\varphi = \pm\pi$), where output at port 1 drops to zero in both the ideal ZIM network (upper) and its waveguide-based implementation (lower). The geometric and electromagnetic parameters are provided in sections 3.2 and 3.4 of Supporting Information, the S-parameter spectra of the waveguide-based implementation are given in section 3.6, and the quantitative analysis of the deviation between the achieved and target S-parameters is provided in section 3.7. These results demonstrate the coherent signal aggregation capabilities of the ZIM network, providing a powerful approach for reflectionless coherent light control that is highly attractive for applications such as modulators, switches, and on-chip signal processors.

## 5. Coherent spatial mode demultiplexer

The fundamental operations demonstrated above, i.e., unicast routing, multicast routing, and coherent beam combining, can be integrated to enable advanced signal-processing functionalities. A representative example is the four-port coherent spatial mode demultiplexer shown in Fig. 5(a), which operates analogously to a Magic-T hybrid junction [61,62]. Here, we directly employ the waveguide-based implementation platform operating in the $TE_{10}$ mode near its cutoff frequency.

The mode demultiplexer can be viewed as a combination of two multicast routers: one routes signals from port 1 to ports 3 and 4, and the other routes signals from port 2 to the same output ports. When ports 1 and 2 are excited coherently, the network is designed to direct common-mode inputs to port 3 and differential-mode inputs to port 4, all without reflection. This functionality requires a scattering matrix:

$$\boldsymbol{S} = \begin{pmatrix} 0 & 0 & \alpha'' & \alpha'' \\ 0 & 0 & \alpha'' & -\alpha'' \\ \alpha'' & \alpha'' & 0 & 0 \\ \alpha'' & -\alpha'' & 0 & 0 \end{pmatrix}, \tag{7}$$

where $\alpha''$ is an arbitrary complex transmission coefficient. Again, applying our inverse design formalism [Eq. (2)] yields analytical expressions for the characteristic parameters:

$$\xi_{11} = \xi_{33} = -i\frac{2\alpha''^2+4\alpha''+1}{2\alpha''^2-1},\ \xi_{22} = \xi_{44} = -i\frac{2\alpha''^2+1}{2\alpha''^2-1},$$
$$\xi_{13} = \xi_{14} = \xi_{23} = -\xi_{24} = i\frac{2\alpha''^2-1}{2\alpha''^2},$$
$$\text{and } \xi_{12} = \xi_{34} \to \infty. \tag{8}$$

From these $\xi_{mn}$ values, the required geometric and electromagnetic parameters of the waveguide-based ZIM network can be directly determined (see Supporting Information, sections 3.2 and 3.5).

As an example, choose $\alpha'' = i/\sqrt{2}$, for which all required $\mu_{mn}$ are purely real, indicating a lossless system. We excite ports 1 and 2 simultaneously with magnetic fields $H_0$ and $e^{i\Delta\varphi}H_0$, respectively. Figure 5(b) presents the simulated normalized output power $P_m$ ($m$ =1,2,3,4) as a function of relative phase difference $\Delta\varphi$. The results show $P_1 = P_2 = 0$, confirming reflectionless operation. Meanwhile, $P_3$ and $P_4$ vary between zero and unity.

Specifically, for in-phase inputs ($\Delta\varphi = 0$), we obtain $P_3 = 1$ and $P_4 = 0$, indicating perfect common-mode routing to port 3, as further verified in the simulated distribution of $H_z/H_0$ in Fig. 5(c). Conversely, for out-of-phase inputs ($\Delta\varphi = \pm\pi$), we obtain $P_3 = 0$ and $P_4 = 1$, corresponding to complete differential-mode routing to port 4, as confirmed by the simulated field distribution in Fig. 5(d). The corresponding S-parameter spectra are provided in section 3.6 of Supporting Information, and the quantitative analysis of the deviation between the achieved and target S-parameters is provided in section 3.7. These results demonstrate the ideal mode sorting performance enabled by the ZIM network without introducing loss or gain.

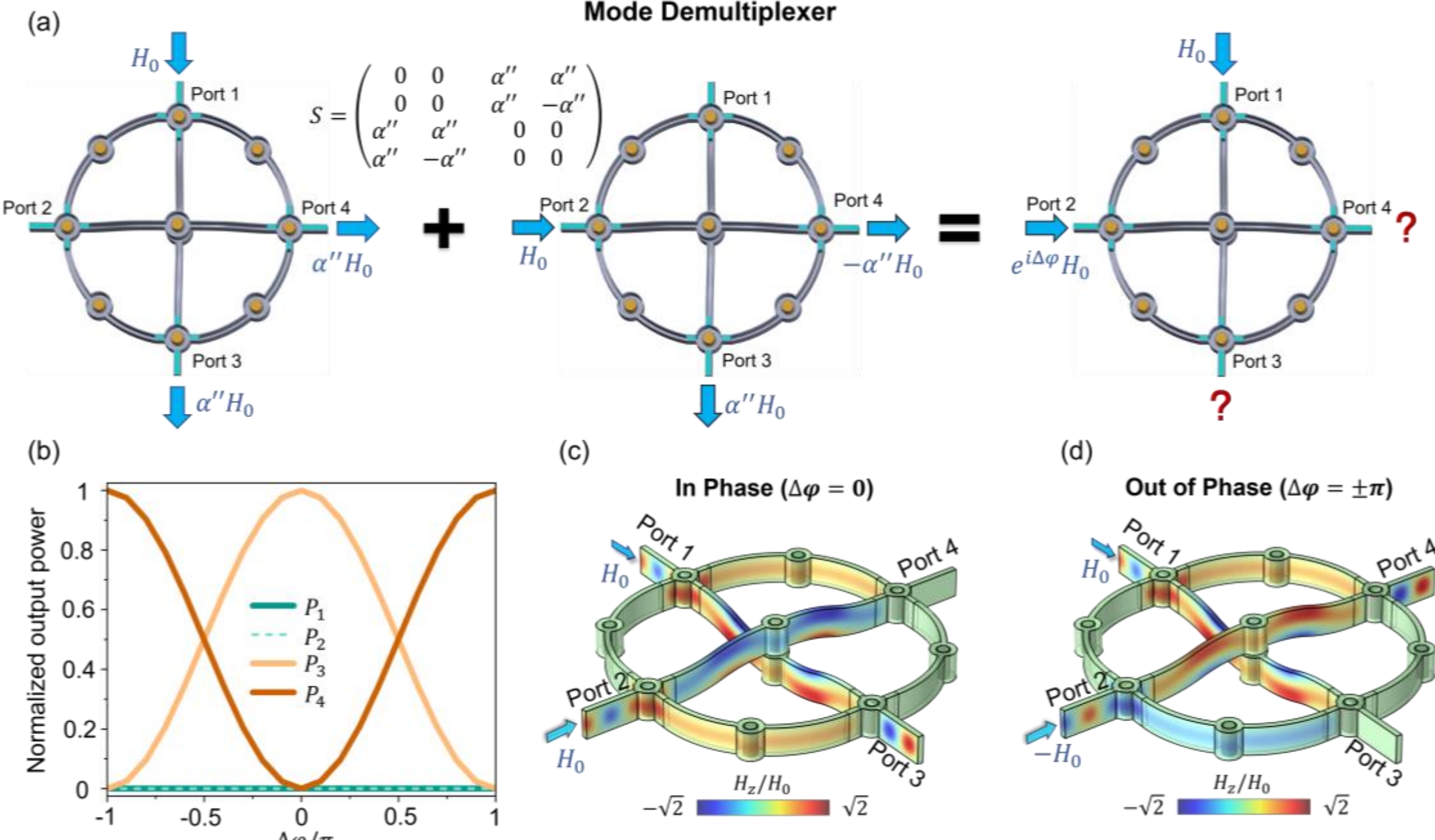

**Figure 5.** A coherent spatial mode demultiplexer. (a) The mode demultiplexer (right) can be viewed as a combination of two multicast routers: one routes signals from port 1 to ports 3 and 4 (left), and the other routes signals from port 2 to the same output ports (middle). Its wave behavior is described by the target scattering matrix $\boldsymbol{S}$ (inset) under coherent excitation of ports 1 and 2. (b) Simulated output power $P_m$ ($m$ =1,2,3,4) as a function of the relative phase difference $\Delta\varphi$ when choosing $\alpha'' = i\sqrt{2}/2$. [(c) and (d)] Simulated distributions of $H_z/H_0$ in the waveguide-based implementation operating at $\omega_c$ for (c) in-phase inputs ($\Delta\varphi = 0$) and (d) out-of-phase inputs ($\Delta\varphi = \pm\pi$).

## 6. Six-port reflectionless routing

Our analytical inverse-design approach based on non-Hermitian ZIM networks is general and can be readily extended to networks with an arbitrary number of ports, thereby enabling more complex routing functionalities. Figure 6 demonstrates an example of six-port reflectionless routing in an ideal ZIM network. We choose a target scattering matrix for this network as:

$$\boldsymbol{S}=\begin{pmatrix} 0 & 0 & 0.5i & 0.5i & 0.5i & -0.5i \\ 0 & 0 & 0.5i & 0.5i & -0.5i & 0.5i \\ 0.5i & 0.5i & 0 & 0 & 0 & 0 \\ 0.5i & 0.5i & 0 & 0 & 0 & 0 \\ 0.5i & -0.5i & 0 & 0 & 0 & 0 \\ -0.5i & 0.5i & 0 & 0 & 0 & 0 \end{pmatrix}. \tag{9}$$

The corresponding characteristic parameters $\xi_{mn}$ derived from Eq. (2), along with the corresponding geometry and permeability $\mu_{mn}$ of the ZIM network, are provided in section 4.1 of Supporting Information.

The scattering matrix in Eq. (9) describes a device that routes a coherent two-port input (ports 1 and 2) to distinct output pairs, depending on the relative phase of the inputs. Figure 6(a) shows the simulated distribution of normalized magnetic field $H_z/H_0$ when in-phase signals (magnetic field $H_0$ at both ports 1 and 2) are applied. We see that the power is routed reflectionlessly and equally to ports 3 and 4, corresponding to common-mode operation. Conversely, for out-of-phase inputs ($H_0$ at port 1, $-H_0$ at port 2), the signal is routed equally to ports 5 and 6, demonstrating the differential-mode behavior [Fig. 6(b)].

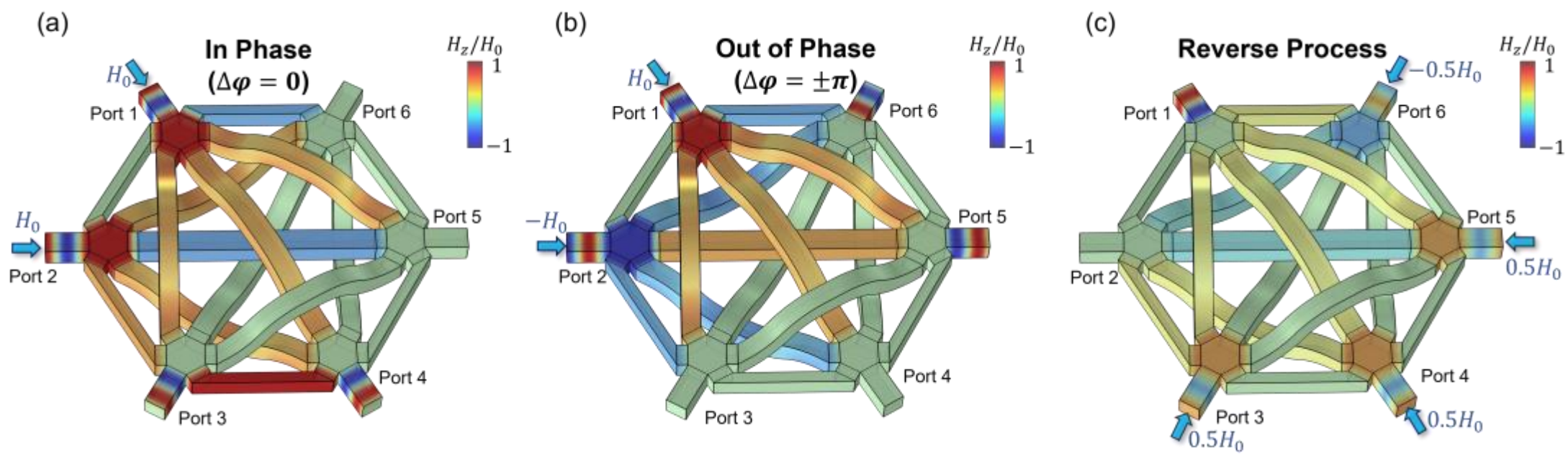

**Figure 6.** A six-port reflectionless router. [(a)-(c)] Simulated distributions of $H_z/H_0$ in an ideal ZIM network under different coherent excitations: (a) in-phase excitation of ports 1 and 2 (both with magnetic field $H_0$), (b) out-of-phase excitation of ports 1 and 2 ($H_0$ at port 1, $-H_0$ at port 2), and (c) simultaneous excitation of ports 3-6 with magnetic fields $0.5H_0$, $0.5H_0$, $0.5H_0$, and $-0.5H_0$, respectively.

Notably, the mode-sorting behavior changes when the device is operated in reverse, that is, when ports 3-6 are coherently excited. Superposing the reversed processes of the two models in Figs. 6(a) and 6(b) leads to constructive interference at port 1 and complete destructive interference at port 2. Consequently, all input power is guided into port 1 without reflection, as confirmed by the simulated field distribution in an ideal ZIM network under the coherent excitation of ports 3-6 with magnetic fields $0.5H_0$, $0.5H_0$, $0.5H_0$, and $-0.5H_0$, respectively [Fig. 6(c)]. The corresponding waveguide-based implementation of this six-port router is provided in sections 4.2 and 4.3 of Supporting Information. These results demonstrate the unprecedented reflectionless mode sorting in this six-port ZIM network, underscoring the power and generality of our analytical inverse-design approach for arbitrary reflectionless optical routing.

## 7. Discussion and conclusion

We note that within the present inverse-design framework, the set of realizable target scattering matrices is subject to two main constraints: 1) the scattering matrix must be symmetric, $S_{mn} = S_{nm}$, due to reciprocity; and 2) the matrix $\boldsymbol{S}+\boldsymbol{I}$ must be non-singular, i.e., $\det(\boldsymbol{S}+\boldsymbol{I}) \neq 0$. Realizing asymmetric scattering matrices would require introducing optical nonreciprocity into ZIMs, e.g., through incorporating magneto-optical materials [63] or nonlinear responses [37], which offer viable routes to extend the framework in future work. The second constraint stems from the inverse-design formula in Eq. (1), which explicitly requires $\boldsymbol{S}+\boldsymbol{I}$ to be non-singular. This limitation is methodological rather than fundamental: a singular $\boldsymbol{S}+\boldsymbol{I}$ does not necessarily imply that the corresponding scattering response is physically unrealizable. For example, special cases such as diagonal scattering matrices with $S_{mm} = -1$ can still be achieved through alternative methods, e.g., by enforcing PMC-like boundary conditions at all nodes.

The proposed ZIM networks operate within a narrow frequency band because ZIMs are intrinsically dispersive, as required by causality and the Kramers-Kronig relations. The specific operating frequency regime is determined by the physical ZIM platform used for implementation. The near-zero permittivity/permeability therefore occurs only around a specific frequency [64], e.g., at the frequency $\omega_c$ in Fig. 3(b). Although this constraint precludes a strictly dispersionless zero-index response, spectral engineering strategies can be employed to broaden the frequency range over which the effective permittivity/permeability remains close to zero, thereby enabling a finite operational bandwidth [65-66].

The practical implementation of the proposed ZIM networks may exhibit sensitivity to fabrication imperfections. A detailed sensitivity analysis is provided section 3.8 of Supporting Information, where Monte Carlo perturbations are introduced to the waveguide-based implementations of the unicast router (Fig. 3), multicast router (Fig. 4), and mode demultiplexer (Fig. 5). We find that, relatively small (sub-percent-level) variations in the dopants' permittivities are required to maintain good performance. By contrast, the characteristic parameter $\xi$ exhibits a greater tolerance, indicating that the device performance is more robust with respect to variations in $\xi$ than in the underlying physical parameters, and that the sensitivity mainly originates from the photonic-doping implementation. In practical implementations, such sensitivity could be alleviated through further parameter optimization[33,34,54-57].

It is noteworthy that, besides the waveguide-based implementation demonstrated above, the proposed ZIM networks can also be realized through alternative platforms. For instance, photonic crystals exhibiting Dirac-like conical dispersions offer a well-established route to realize effective ZIMs in both two and three dimensions, spanning regimes from microwave to optical frequencies [46, 67-72]. Moreover, the possibility of doping such photonic-crystal-based ZIMs to achieve desired complex effective parameters (e.g., $\mu_{mn}$ in this work) has been demonstrated both theoretically and experimentally [34, 36, 40], offering a viable path toward practical realization of the proposed ZIM networks for arbitrary reflectionless optical routing.

Finally, regarding the physical layout of the proposed networks, small-scale configurations (e.g., $N \leq 4$) can indeed be implemented using a strictly planar geometry without physical waveguide crossings, as detailed in Section 5 of the Supporting Information. However, for larger routing networks ($N > 4$), the fully connected topology becomes inherently non-planar, necessitating the three-dimensional overpass architecture adopted in our current design to prevent unwanted interference. Looking forward, recent advances in nonlocal photonic crystals may provide a promising alternative to resolve these fundamental crossing issues, potentially enabling densely connected routing networks in a strictly two-dimensional planar platform [73,74].

In summary, we have developed an analytical inverse-design approach for arbitrary reflectionless optical routing using $N$-port non-Hermitian ZIM networks. The core of the approach is an algebraic mapping that deterministically converts target scattering matrix into the physical parameters of the network, thereby eliminating the need for iterative numerical optimization. This mapping enables the systematic design of arbitrary reflectionless routing devices. We have demonstrated the versatility of the non-Hermitian ZIM networks by designing a range of functional devices, from unicast and multicast routers with full amplitude and phase control to coherent beam combiners and spatial mode demultiplexers, in four-port and six-port networks. This work establishes a direct analytical pathway to high-performance reflectionless optical routers, with potential implications in advanced modulators, switches, and on-chip signal processors.

**Supporting Information**

Supporting Information is available from the Wiley Online Library or from the author.

**Acknowledgments.**

This work was supported by National Natural Science Foundation of China (Grant Nos. 12104191, 12374293); Natural Science Research of Jiangsu Higher Education Institutions of China (Grant No. 21KJB140006); Natural Science Foundation of Jiangsu Province (Grant No. BK20233001).

**Conflict of Interest**

The authors declare no conflict of interest.

**Data Availability Statement**

The data that support the findings of this study are available from the corresponding author upon reasonable request.

Supporting Information

# Arbitrary Reflectionless Optical Routing via Non-Hermitian Zero-Index Networks

*Yongxing Wang*, Zehui Du, Zhenshuo Xu, Pei Xiao, Jizi Lin, Yufeng Zhang, and Jie Luo**

Y. Wang, Z. Du, Z. Xu, P. Xiao, J. Lin,

Zhangjiagang Campus, Jiangsu University of Science and Technology, Zhangjiagang 215600, China

Email: 201900000107@just.edu.cn

Y. Wang, P. Xiao, J. Lin, Y. Zhang

Department of Physics, Jiangsu University of Science and Technology Suzhou Institute of Technology, Zhangjiagang 215600, China

J. Luo

School of Physical Science and Technology, Soochow University, Suzhou 215006, China

Email: luojie@suda.edu.cn

J. Luo

Jiangsu Physical Science Research Center, Nanjing 210093, China

## 1. Derivation of the analytical inverse design formula

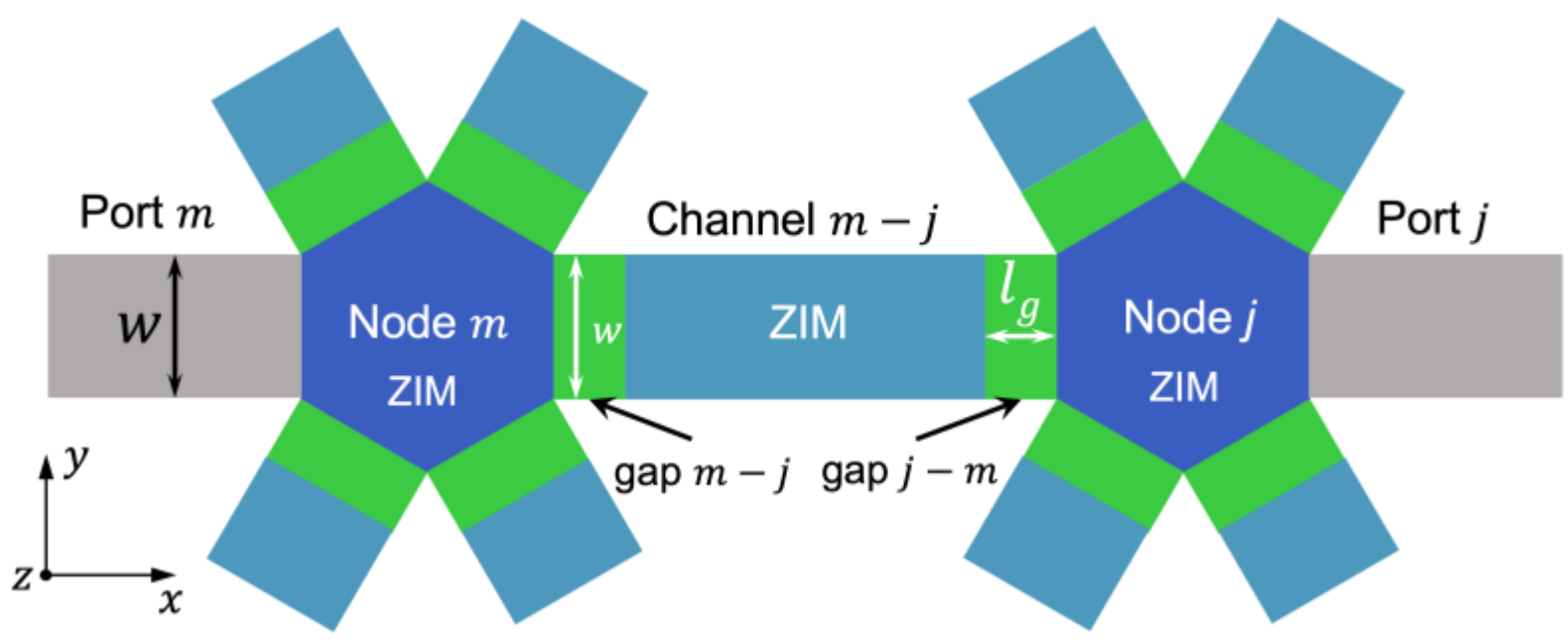

**Figure S1.** Details and geometric parameters of two connected nodes in the ZIM network. This schematic illustrates two adjacent nodes, node $m$ and node $j$, connected by a channel $m$-$j$.

We consider that the $n$-th port is excited by a transverse-magnetic (TM) polarized plane wave with amplitude $H_0$ and angular frequency $\omega$. Assuming a time variation term of $\mathrm{e}^{-i\omega t}$, which is omitted henceforth for brevity, the magnetic field within port $m$ (or $j$) can be expressed as

$$\mathbf{H}_{m(j)} = H_0\left(\delta_{m(j)n}e^{i\mathbf{k}_0\cdot\mathbf{r}} + S_{m(j)n}e^{-i\mathbf{k}_0\cdot\mathbf{r}}\right)\mathbf{e}_z, \tag{S1}$$

where $\mathbf{k}_0 = (\omega/c)\,\mathbf{e}_{\mathbf{k}_0}$ is wave vector in the port with the unit vector $\mathbf{e}_{\mathbf{k}_0}$ vertical pointing to the connecting node; $S_{m(j)n}$ is the S-parameters defined as the ratio of the output magnetic field within port $m$ (or $j$) to the input magnetic field within port $n$. Similarly, the magnetic field within gap $m$-$j$ (or $j$-$m$) is

$$\mathbf{H}_{mj(jm)} = H_0\left(a_{mj(jm)}e^{i\mathbf{k}_g\cdot\mathbf{r}} + b_{mj(jm)}e^{-i\mathbf{k}_g\cdot\mathbf{r}}\right)\mathbf{e}_z, \tag{S2}$$

where $\mathbf{k}_g = \sqrt{\varepsilon_g}\mathbf{k}_0$ is wave vector in gaps, $\varepsilon_g$ is the permittivity of gaps; $H_0 a_{mj(jm)}$ and $H_0 b_{mj(jm)}$ are magnetic field complex amplitude of backward and forward waves within gap $m$-$j$ (or $j$-$m$). By utilizing the Ampère-Maxwell equation, the electric fields within port $m$ (or $j$) and gap $m$-$j$ (or $j$-$m$) are

$$\mathbf{E}_{m(j)} = z_p z_0 H_0\left(\delta_{m(j)n}e^{i\mathbf{k}_0\cdot\mathbf{r}} - S_{m(j)n}e^{-i\mathbf{k}_0\cdot\mathbf{r}}\right)\mathbf{e}_{\mathbf{k}_0}\times\mathbf{e}_z, \tag{S3}$$

$$\text{and}\quad \mathbf{E}_{mj(jm)} = z_g z_0 H_0\left(a_{mj(jm)}e^{i\mathbf{k}_g\cdot\mathbf{r}} - b_{mj(jm)}e^{-i\mathbf{k}_g\cdot\mathbf{r}}\right)\mathbf{e}_{\mathbf{k}_0}\times\mathbf{e}_z, \tag{S4}$$

where $z_0$ is the impendence in vacuum; $z_p$ and $z_g$ are the relative impendence of ports and gaps. For the sake of simplicity in our analysis, we set $z_p = z_g$. Considering the uniform magnetic field within a two-dimensional (2D) zero-index material (ZIM), we apply continuity boundary conditions at port-node, node-gap and gap-channel interfaces and apply the integral form of the Faraday-Maxwell equation to the channel $m$-$j$, we obtain the following relation:

$$\begin{pmatrix} 1 & 1 & 0 & 0 \\ e^{ik_g l_g} & e^{-ik_g l_g} & -e^{ik_g l_g} & -e^{-ik_g l_g} \\ 0 & 0 & 1 & 1 \\ e^{ik_g l_g}\left(1+i\xi_{mj}\right) & -e^{-ik_g l_g}\left(1-i\xi_{mj}\right) & e^{ik_g l_g} & e^{-ik_g l_g} \end{pmatrix} \begin{pmatrix} a_{mj} \\ b_{mj} \\ a_{jm} \\ b_{jm} \end{pmatrix} = \begin{pmatrix} S_{mn}+\delta_{mn} \\ 0 \\ S_{jn}+\delta_{jn} \\ 0 \end{pmatrix}, \quad \text{(S5)}$$

where $\xi_{mj} = \xi_{jm} \equiv k_0\, \mu_{mj} A_{mj}/z_p w$. Then, applying the integral form of the Faraday-Maxwell equation to the node $m$ yields

$$-(\delta_{mn} - S_{mn}) + \sum_{\substack{j=1 \\ j\neq m}}^{N} \left(a_{mj} - b_{mj}\right) = i\xi_{mm}\left(a_{mj} + b_{mj}\right). \quad \text{(S6)}$$

Solving the Eq. (S5) and substituting it into Eq. (S6) yields

$$\sum_{\substack{j=1 \\ j\neq m}}^{N} i\,\frac{2(S_{jn}+\delta_{jn})+(S_{mn}+\delta_{mn})[\xi_{jm}\sin(2k_g l_g)-2\cos(2k_g l_g)]}{2\sin(k_g l_g)[\xi_{jm}\sin(k_g l_g)-2\cos(k_g l_g)]} = (i\xi_{mm}-1)(S_{mn}+\delta_{mn}) + 2\delta_{mn}. \quad \text{(S7)}$$

Considering the reciprocity that enforces $S_{mn} = S_{nm}$, this set of relations can be elegantly organized into the following compact matrix equation:

$$(\boldsymbol{S} + \boldsymbol{I})\boldsymbol{W} = -2i\boldsymbol{I}, \quad \text{(S8)}$$

where $\boldsymbol{I}$ is the $N$-order identity matrix. The elements of the auxiliary matrix $\boldsymbol{W}$ are given by:

$$w_{mn} = \begin{cases} -\left(\xi_{mm} + i\right) + \sum_{\substack{j=1 \\ j\neq m}}^{N} \dfrac{\left[\xi_{mj}\sin\left(2k_g l_g\right)-2\cos\left(2k_g l_g\right)\right]}{2\sin\left(k_g l_g\right)\left[\xi_{mj}\sin\left(k_g l_g\right)-2\cos\left(k_g l_g\right)\right]}, & (m = n) \\ \dfrac{1}{\sin\left(k_g l_g\right)\left[\xi_{mn}\sin\left(k_g l_g\right)-2\cos\left(k_g l_g\right)\right]}, & (m \neq n) \end{cases} \quad \text{(S9)}$$

Finally, by solving the matrix Eq. (S8) and combining with Eq. (S9), we can obtain the required geometric and electromagnetic parameters of each ZIM component (node and channel) from the target scattering matrix $\boldsymbol{S}$ via:

$$\xi_{nn} = \frac{-2i\sum_{i=1}^{N}(1-2\delta_{in})\mathrm{C}_{ni}(\boldsymbol{S}+\boldsymbol{I})}{\det(\boldsymbol{S}+\boldsymbol{I})} + (N-1)\cot\left(k_g l_g\right) - i \quad \text{(S10a)}$$

$$\xi_{mn} = \frac{i\det(\boldsymbol{S}+\boldsymbol{I})}{2\sin^2\left(k_g l_g\right)\mathrm{C}_{mn}(\boldsymbol{S}+\boldsymbol{I})} + 2\cot\left(k_g l_g\right), (m \neq n) \quad \text{(S10b)}$$

where $\det(\boldsymbol{S}+\boldsymbol{I})$ and $\mathrm{C}_{mn}(\boldsymbol{S}+\boldsymbol{I})$ signify the determinant and the $(m,\ n)$-cofactor of the matrix $\boldsymbol{S}+\boldsymbol{I}$, respectively.

## 2. Four-port ideal ZIM model (Figs. 2, 4, 5)

### 2. 1. General logic and geometry (four-port)

This section provides the geometric and electromagnetic parameters for the ideal ZIM models presented in Figs. 2 and 4 in Main Text. The common geometric parameters are $l_g = 0.25\lambda_0$ and $w = h = 0.5\lambda_0$ for all ideal ZIM models in Main Text.

These models use a three-dimensional (3D) architecture where diagonal channels (channels 1-3 and 2-4) bend out-of-plane, forming an overpass-like structure. Since the relation for characteristic parameters $\xi_{mn} = k_0\mu_{mn}A_{mn}/z_p w$ derived in section 1 assumes a planar 2D geometry, the spatial curvature and path elongation in the 3D model inevitably introduce minor numerical deviations for these bent channels. Given that each target $\xi_{mn}$ is pre-determined by theoretically [according to Eqs. (S10a) and (S10b)], this refinement requires only modest adjustments to the corresponding parameter for each component. To clearly present these parameters, the following Tables S1 and S2 show the progression for each 'ZIM component'. The tables list the target 'Calculated $\xi_{mn}$', which is calculated from the analytical expressions using the S-parameters. From this target value, we obtain both the 'Initial $\mu_{mn}$' based on the 2D analytical model, and the 'Refined $\mu_{mn}$', which represents the final numerically optimized value used in the 3D simulation. For planar units, including nodes and adjacent channels, the 'Refined $\mu_{mn}$' column is marked "same as initial" as its value is identical to the initial one. For perfect magnetic conductor (PMC) components ($\xi = \infty$), both $\mu_{mn}$ columns are marked "N/A (PMC)".

The four-port network consists of 4 nodes and 6 channels. The area $A_{mn}$ is defined with the following symmetries: $A_{11} = A_{22} = A_{33} = A_{44} = 0.25\lambda_0^2$ for ZIM nodes; $A_{12} = A_{23} = A_{34} = A_{14} = 0.8\lambda_0^2$ and $A_{13} = A_{24} = 1.6436\lambda_0^2$ for ZIM channels. In our 3D model, the channels 1-3 and 2-4 are bent out-of-plane, while all nodes and other channels are planar.

### 2.2. Parameters for the dual-channel unicast router (Fig. 2)

For the ZIM-network-based unicast router with $\alpha = i$ and $\beta = 2$ [Figs. 2(c)-2(e) in Main Text], the calculated $\xi_{mn}$, initial $\mu_{mn}$, and refined $\mu_{mn}$ used in simulations are listed in Table S1.

**Table S1: Ideal ZIM parameters for the unicast router.**

| ZIM units | Calculated $\xi_{mn}$ | Initial $\mu_{mn}$ | Refined $\mu_{mn}$ |
|---|---|---|---|
| Node 1 | $-1$ | $-0.3183$ | Same as Initial |
| Node 2 | $-1$ | $-0.3183$ | Same as Initial |
| Node 3 | $-3i$ | $-0.9549i$ | Same as Initial |
| Node 4 | $-3i$ | $-0.9549i$ | Same as Initial |
| Channel 1-2 | $-1$ | $-0.0995$ | Same as Initial |
| Channel 1-3 | $\infty$ | $\infty$ | N/A (PMC) |
| Channel 1-4 | $\infty$ | $\infty$ | N/A (PMC) |
| Channel 2-3 | $\infty$ | $\infty$ | N/A (PMC) |
| Channel 2-4 | $\infty$ | $\infty$ | N/A (PMC) |
| Channel 3-4 | $0.75i$ | $0.0746i$ | Same as Initial |

### 2.3. Multicast router (Fig. 4)

For the ZIM-network-based multicast router with a target scattering matrix given by Eq. (5) for $\alpha' = -i\sqrt{1/3}$ and $\beta' = -i\sqrt{2/3}$ [Figs. 4(b)-(e) in Main Text], the calculated $\xi_{mn}$, initial $\mu_{mn}$, and refined $\mu_{mn}$ used in simulations are listed in Table S2.

**Table S2: Ideal ZIM parameters for the multicast router**

| ZIM units | Calculated $\xi_{mn}$ | Initial $\mu_{mn}$ | Refined $\mu_{mn}$ |
|---|---|---|---|
| Node 1 | $1.3938$ | $0.4437$ | Same as Initial |
| Node 2 | $0.5774 + 1.1381i$ | $0.1838 + 0.3623i$ | Same as Initial |
| Node 3 | $0.8165 + 0.8047i$ | $0.2599 + 0.2562i$ | Same as Initial |
| Node 4 | $i$ | $0.3183i$ | Same as Initial |
| Channel 1-2 | $1.7321$ | $0.1723$ | Same as Initial |
| Channel 1-3 | $1.2247$ | $0.0593$ | $0.0596$ |
| Channel 1-4 | $\infty$ | $\infty$ | N/A (PMC) |
| Channel 2-3 | $-2.1213i$ | $-0.2110i$ | Same as Initial |
| Channel 2-4 | $\infty$ | $\infty$ | N/A (PMC) |
| Channel 3-4 | $\infty$ | $\infty$ | N/A (PMC) |

## 3. Four-port waveguide implementation (Figs. 3, 4, 5)

### 3.1. Waveguide platform and photonic doping model

Our practical implementation is based on perfect electric conductor (PEC) waveguides operating at the $TE_{10}$-mode cutoff frequency $\omega_c$. The effective permittivity for the $TE_{10}$ mode in such a waveguide is generally described by the dispersion relation [1-2]:

$$\varepsilon^{\text{eff}} = \varepsilon_f - \frac{\omega_c^2}{\omega^2} \tag{S11}$$

where $\varepsilon_f$ is the relative permittivity of the material filling the waveguide, and $\omega_c = c\pi/w$ is the cutoff frequency of the air-filled waveguide of the same dimensions. This single formula explains the behavior of our entire platform at the operating frequency $\omega = \omega_c$. For the ZIM 'nodes' and 'channels', which are air-filled ($\varepsilon_f = 1$), the effective permittivity becomes $\varepsilon^{\text{eff}}(\omega_c) = 0$. This creates a ZIM background with $\varepsilon^{\text{eff}}(\omega_c) = 0$. For the input/output ports and coupling gaps correspond to the Teflon-filled ($\varepsilon_f = 2.1$) waveguide regions, with effective permittivity $\varepsilon^{\text{eff}}(\omega_c) = 1.1$ and the corresponding effective relative impedance $z^{\text{eff}}(\omega_c) = \sqrt{1/1.1}$.

The physical realization of each network component is based on the photonic doping theory [3-4]. To determine the required dopant properties for each component, we substitute the relation for characteristic parameters $\xi_{mn} = k_0\mu_{mn}A_{mn}/z_p w$, into the doping model, which relates the effective permeability $\mu_{mn}$ to the dopant's radius $R_{d,mn}$ and its effective permittivity $\varepsilon_{d,mn}^{\text{eff}}$ within the 2D analytical model. This yields the following expression:

$$\xi_{mn} = \frac{k_0(A_{mn} - \pi R_{d,mn}^2)}{z_p w} + \frac{2\pi R_{d,mn} J_1(\sqrt{\varepsilon_{d,mn}^{\text{eff}}} k_0 R_{d,mn})}{z_p w \sqrt{\varepsilon_{d,mn}^{\text{eff}}}\, J_0\left(\sqrt{\varepsilon_{d,mn}^{\text{eff}}} k_0 R_{d,mn}\right)}, \tag{S12}$$

where $J_0$ and $J_1$ respectively denote 0-order and 1-order Bessel function of the first kind. The final permittivity of the dopant material $\varepsilon_{d,mn}^{\text{eff}}$ used in the 3D waveguide implementation is obtained via $\varepsilon_{d,mn} = \varepsilon_{d,mn}^{\text{eff}} + 1$. This relation ensures the target response is correctly mapped to the physical waveguide platform at the cutoff frequency, consistent with Eq. (S11).

The relation $\xi_{mn} = k_0\mu_{mn}A_{mn}/z_p w$ is based on an ideal 2D model, which does not account for the complex 3D out-of-plane geometry of the network. Therefore, these initial dopant permittivity values $\varepsilon_{d,mn}$ are further refined in the 3D waveguide simulation models. Tables (S3)-(S5) present both the initial and refined values used in simulations.

### 3.2. Common geometric parameters (four-port)

The four-port waveguide-based implementation in Figs. 3-5 of Main Text has some common geometric parameters: waveguide height $h = 0.5\lambda_0$, waveguide width $w = 0.1\lambda_0$, Teflon-filled gap length $l_g = 0.25\lambda_0/\sqrt{1.1}$, dopant radius $R_{d,mn} = 0.1\lambda_0$, node areas $A_{11} = A_{22} = A_{33} = A_{44} = 0.1631\lambda_0^2$, and channel areas $A_{12} = A_{23} = A_{34} = A_{14} = 0.4206\lambda_0^2$, $A_{13} = A_{24} = 0.5126\lambda_0^2$.

### 3.3. Unicast router dopant parameters (Fig. 3)

The parameters for the waveguide-based unicast router ($\alpha = i, \beta = 2$) are listed in Table S3. The 'Initial $\varepsilon_{d,mn}$' is calculated from the doping model (Eq. S12) with the $TE_{10}$ dispersion relation (Eq. S11). The 'Refined $\varepsilon_{d,mn}$' is optimized in full-wave simulations.

**Table S3: Dopant permittivities for the waveguide based unicast router**

| ZIM units | Target $\xi_{mn}$ | Initial $\varepsilon_{d,mn}$ | Refined $\varepsilon_{d,mn}$ |
|---|---|---|---|
| Node 1 | $-1$ | 17.6646 | 17.6581 |
| Node 2 | $-1$ | 17.6646 | 17.6581 |
| Node 3 | $-3i$ | $17.6735 - 0.6447i$ | $17.6685 - 0.6424i$ |
| Node 4 | $-3i$ | $17.6735 - 0.6447i$ | $17.6685 - 0.6424i$ |
| Channel 1-2 | $-1$ | 16.4154 | 16.4113 |
| Channel 1-3 | $\infty$ | 15.6490 | 15.6490 |
| Channel 1-4 | $\infty$ | 15.6490 | 15.6490 |
| Channel 2-3 | $\infty$ | 15.6490 | 15.6490 |
| Channel 2-4 | $\infty$ | 15.6490 | 15.6490 |
| Channel 3-4 | $0.75i$ | $16.4439 + 0.0226i$ | $16.4398 + 0.0226i$ |

### 3.4. Multicast router dopant parameters [Fig. 4(d), 4(e)]

The parameters for the waveguide-based multicast router ($\alpha' = -i\sqrt{1/3}, \beta' = -i\sqrt{2/3}$) are presented in Table S4.

**Table S4: Dopant permittivities for the waveguide based multicast router**

| ZIM units | Target $\xi_{mn}$ | Initial $\varepsilon_{d,mn}$ | Refined $\varepsilon_{d,mn}$ |
|---|---|---|---|
| Node 1 | 1.3938 | 18.2654 | 18.2550 |
| Node 2 | $0.5774 + 1.1381i$ | $17.9854 + 0.3005i$ | $17.9760 + 0.2991i$ |
| Node 3 | $0.8165 + 0.8047i$ | $18.0689 + 0.2262i$ | $18.0590 + 0.2256i$ |
| Node 4 | $i$ | $17.8537 + 0.2340i$ | $17.8450 + 0.2326i$ |
| Channel 1-2 | 1.7321 | 16.5004 | 16.4961 |
| Channel 1-3 | 1.2247 | 16.3212 | 16.3056 |
| Channel 1-4 | $\infty$ | 15.6490 | 15.6490 |
| Channel 2-3 | $-2.1213i$ | $16.4394 - 0.0636i$ | $16.4352 - 0.0635i$ |
| Channel 2-4 | $\infty$ | 15.6490 | 15.6490 |
| Channel 3-4 | $\infty$ | 15.6490 | 15.6490 |

### 3.5. Coherent mode demultiplexer dopant parameters

The parameters for the waveguide-based mode demultiplexer ($\alpha'' = i/\sqrt{2}$) are presented in Table S5.

**Table S5: Dopant permittivities for the waveguide based mode demultiplexer**

| ZIM units | Target $\xi_{mn}$ | Initial $\varepsilon_{d,mn}$ | Refined $\varepsilon_{d,mn}$ |
|---|---|---|---|
| Node 1 | $-1.4142$ | 17.5875 | 17.5815 |
| Node 2 | 0 | 17.8785 | 17.8700 |
| Node 3 | $-1.4142$ | 17.5875 | 17.5815 |
| Node 4 | 0 | 17.8785 | 17.8700 |
| Channel 1-2 | $\infty$ | 15.6490 | 15.6490 |
| Channel 1-3 | $-1.4142$ | 16.2688 | 16.2508 |
| Channel 1-4 | $-1.4142$ | 16.4040 | 16.3999 |
| Channel 2-3 | $-1.4142$ | 16.4040 | 16.3999 |
| Channel 2-4 | 1.4142 | 16.3253 | 16.3099 |
| Channel 3-4 | $\infty$ | 15.6490 | 15.6490 |

### 3.6. S-Parameter spectra for waveguide implementations

Here, we present the simulated S-parameters for the waveguide-based implementations of the multicast router and coherent mode demultiplexers, as shown in Fig. S2. Figure S2(a) displays the S-parameter spectra for the waveguide-based multicast router (the model in Fig. 4 of Main

Text). At the TE$_{10}$ cutoff frequency ($\omega = \omega_c$), the device achieves the desired performance: all reflections $|S_{11}|$, $|S_{22}|$, $|S_{33}|$, $|S_{44}|$ and undesired couplings $|S_{14}|$, $|S_{23}|$, $|S_{24}|$, $|S_{34}|$ are suppressed to be near-zero. Simultaneously, the key transmission paths converge to their theoretical target magnitudes, with $|S_{12}|=|S_{21}|\approx 0.577$ and $|S_{13}|=|S_{31}|\approx 0.816$.

Figure S2(b) shows the S-parameter spectra for the waveguide-based coherent mode demultiplexer (the model in Fig. 5 of Main Text). At the target frequency $\omega_c$, the simulation results confirm the desired performance: all reflections $|S_{11}|$, $|S_{22}|$, $|S_{33}|$, $|S_{44}|$ and undesired inter-port couplings $|S_{12}|$, $|S_{34}|$ are suppressed to near-zero. Simultaneously, the four mode-sorting transmission paths $|S_{13}|$, $|S_{14}|$, $|S_{23}|$, $|S_{24}|$ all converge to the theoretical target magnitude of $|S_{mn}|\approx 0.707$.

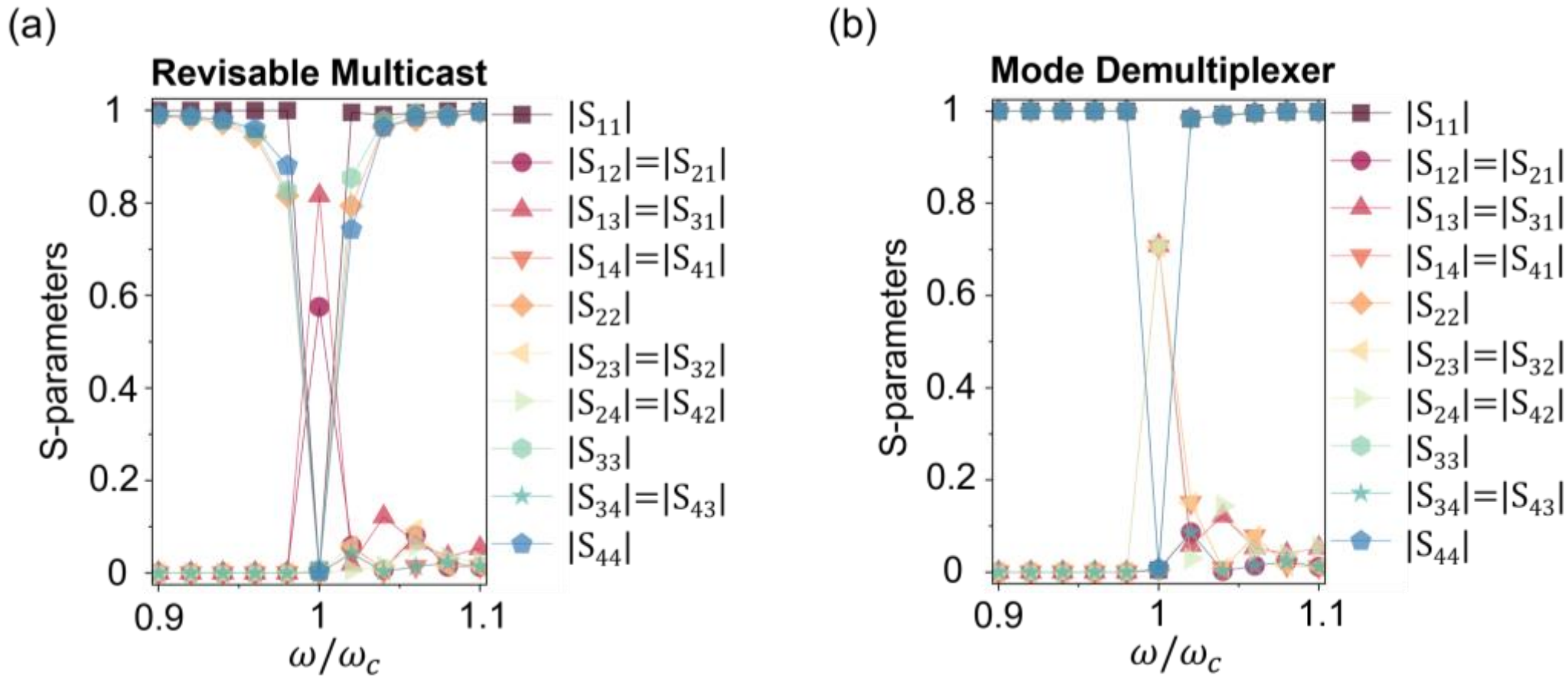

**Figure S2.** S-parameter spectra for four-port waveguide-based implementations. (a) S-parameter spectra for the waveguide-based multicast router, with parameters in presented Table S5, corresponding to the model in Fig. 4 of Main Text. (b) S-parameter spectra for the waveguide-based coherent mode demultiplexer, with parameters in presented Table S6, corresponding to the model in Fig. 5 of Main Text.

### 3.7. Deviations between the achieved and target scattering matrices in waveguide-implemented ZIM networks

Table S6 presents the deviations between the achieved and target S-parameters for the waveguide implementations of the unicast router (Fig. 3), multicast router (Fig. 4), and mode demultiplexer (Fig. 5). We see that the deviations are generally on the order of $10^{-6} \sim 10^{-2}$. To further quantify the overall deviation, we compute the Frobenius norm of the error matrix, defined as $\|\boldsymbol{S}-\boldsymbol{S}^{\text{target}}\|_F = \sqrt{\sum_{m=1}^{4}\sum_{n=1}^{4}\left(S_{mn}-S_{mn}^{\text{target}}\right)^2}$, which characterize the global error magnitude. The values listed in the last row of Table S6 are all small, demonstrating the excellent agreement between the achieved and target scattering matrices for all three devices.

**Table S6: Deviations between the achieved and target S-parameters**

| Deviation | Unicast Router | Multicast Router | Mode Demultiplexer |
|---|---|---|---|
| $\left\|S_{11}-S_{11}^{\text{target}}\right\|$ | $1.78\times10^{-3}$ | $9.75\times10^{-3}$ | $6.12\times10^{-3}$ |
| $\left\|S_{12}-S_{12}^{\text{target}}\right\|$ | $4.51\times10^{-3}$ | $5.57\times10^{-3}$ | $6.90\times10^{-3}$ |
| $\left\|S_{13}-S_{13}^{\text{target}}\right\|$ | $6.34\times10^{-6}$ | $1.10\times10^{-2}$ | $3.66\times10^{-3}$ |
| $\left\|S_{14}-S_{14}^{\text{target}}\right\|$ | $5.93\times10^{-6}$ | $3.54\times10^{-6}$ | $5.46\times10^{-3}$ |
| $\left\|S_{22}-S_{22}^{\text{target}}\right\|$ | $1.78\times10^{-3}$ | $1.37\times10^{-3}$ | $7.25\times10^{-3}$ |
| $\left\|S_{23}-S_{23}^{\text{target}}\right\|$ | $6.50\times10^{-6}$ | $5.21\times10^{-3}$ | $5.76\times10^{-3}$ |
| $\left\|S_{24}-S_{24}^{\text{target}}\right\|$ | $5.69\times10^{-6}$ | $3.23\times10^{-6}$ | $7.61\times10^{-3}$ |
| $\left\|S_{33}-S_{33}^{\text{target}}\right\|$ | $3.26\times10^{-2}$ | $1.25\times10^{-2}$ | $6.37\times10^{-3}$ |
| $\left\|S_{34}-S_{34}^{\text{target}}\right\|$ | $3.15\times10^{-2}$ | $2.16\times10^{-6}$ | $6.68\times10^{-3}$ |
| $\left\|S_{44}-S_{44}^{\text{target}}\right\|$ | $3.25\times10^{-2}$ | $3.36\times10^{-3}$ | $7.44\times10^{-3}$ |
| $\|\boldsymbol{S}-\boldsymbol{S}^{\text{target}}\|_F$ | $6.44\times10^{-2}$ | $2.47\times10^{-2}$ | $2.53\times10^{-2}$ |

### 3.8. Sensitivity analysis for waveguide-based implementations

To evaluate the sensitivity of the proposed ZIM networks to fabrication imperfections, we perform a Monte Carlo analysis by introducing random perturbations to the dopant's relative permittivity $\varepsilon_{d,mn}$ in the waveguide-based implementations of the unicast router, multicast router, and mode demultiplexer, corresponding to the models in Figs. 3-5, respectively. For each parameter set, 2000 random realizations are sampled to obtain statistically stable results.

Figures S3(a)-S3(c) show the normalized deviation $|\Delta S_{mn}/S_{\max}|$ under perturbations in $\varepsilon_{d,mn}$ for the three waveguide-based implementations. Here, $\Delta S_{mn}$ is the difference between the achieved and target S-parameters, and $S_{\max}$ is the maximum magnitude among the target S-parameter elements. $\left(\delta\varepsilon_{d,mn}\right)_{\max}$ denotes the maximum relative variation in $\varepsilon_{d,mn}$. Both the maximum and average deviations between achieved and target scattering matrices over all 16 S-parameter elements are presented. We find that relatively small (sub-percent-level)

variations in the dopants' permittivities are required to maintain good performance.

To further clarify the origin of this sensitivity, we perform Monte Carlo perturbations on the characteristic parameter $\xi_{mn}$, which encapsulates both geometric and electromagnetic parameters [Figs. S3(d)-S3(f)]. We observe that the achieved S-parameters remain in good agreement with the target values under moderate (few-percent-level) variations in $\xi_{mn}$. This indicates that the observed sensitivity mainly originates from the mapping between $\varepsilon_{d,mn}$ and $\xi_{mn}$ in the photonic-doping implementation rather than the inverse-design framework itself. Such sensitivity could be alleviated through further parameter optimization, as demonstrated in experimentally realized photonic-doping-based ZIM systems.

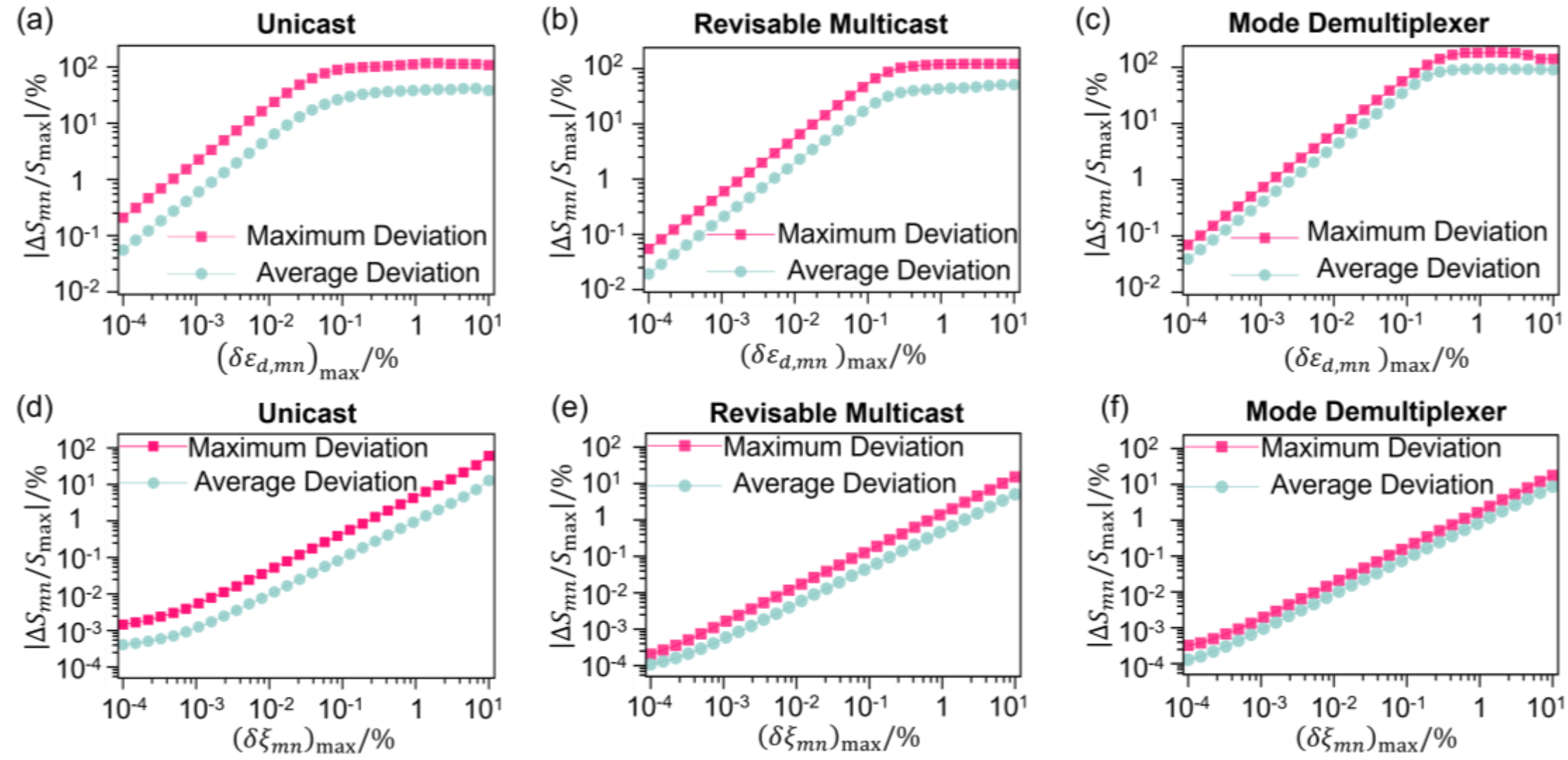

**Figure S3.** Sensitivity analysis based on Monte Carlo perturbations. [(a)-(c)] Deviations of the S-parameters for the (a) unicast router, (b) reversible multicast router, and (c) mode demultiplexer under random perturbations of the dopant's relative permittivity $\varepsilon_{d,mn}$. [(d)-(f)] Corresponding results under perturbations of the characteristic parameter $\xi_{mn}$. Each panel shows both the maximum and average deviations between the achieved and target values across all 16 S-parameter elements.

## 4. Six-port mode sorter (Fig. 6)

### 4.1. Ideal ZIM model (six-port)

**Geometry:** The six-port ideal ZIM model adopts the same parameters as the four-port model: $h = w = 0.5\lambda_0$ and $l_g = 0.25\lambda_0$. The network consists of 6 nodes and 15 channels. The areas for the ZIM nodes are set as $A_{11} = A_{22} = A_{33} = A_{44} = A_{55} = A_{66} = 0.6495\lambda_0^2$. For the ZIM channels, the areas are set as follows: $A_{12} = A_{23} = A_{34} = A_{45} = A_{56} = A_{16} = 0.6793\lambda_0^2$, $A_{13} = A_{24} = A_{35} = A_{46} = A_{51} = A_{62} = 2.1848\lambda_0^2$, $A_{14} = A_{36} = 2.8328\lambda_0^2$, and $A_{25} = 2.7170\lambda_0^2$. To avoid intersections, channels 1-3, 2-4, 3-5, 4-6,5-1, 6-2,1-4, and 3-6 are constructed in an out-of-plane configuration, analogous to an overpass structure.

**Parameters:** The calculated $\xi_{mn}$, initial $\mu_{mn}$, and refined $\mu_{mn}$ used in simulations are presented in Table S7.

**Table S7. Ideal ZIM parameters for the six-port mode sorter**

| ZIM units | Calculated $\xi_{mn}$ | Initial $\mu_{mn}$ | Refined $\mu_{mn}$ |
|---|---|---|---|
| Node 1 | $-1$ | $-0.1225$ | Same as Initial |
| Node 2 | $-1$ | $-0.1225$ | Same as Initial |
| Node 3 | $-1+i$ | $-0.12251 + 0.1225i$ | Same as Initial |
| Node 4 | $-1+i$ | $-0.12251 + 0.1225i$ | Same as Initial |
| Node 5 | $0$ | $0$ | Same as Initial |
| Node 6 | $0$ | $0$ | Same as Initial |
| Channel 1-2 | $\infty$ | $\infty$ | N/A (PMC) |
| Channel 1-3 | $-2$ | $-0.0729$ | $-0.0733$ |
| Channel 1-4 | $-2$ | $-0.0585$ | $-0.0586$ |
| Channel 1-5 | $-2$ | $-0.0729$ | $-0.0733$ |
| Channel 1-6 | $2$ | $0.2343$ | Same as Initial |
| Channel 2-3 | $-2$ | $-0.2343$ | Same as Initial |
| Channel 2-4 | $-2$ | $-0.0729$ | $-0.0733$ |
| Channel 2-5 | $2$ | $0.0563$ | Same as Initial |
| Channel 2-6 | $-2$ | $-0.0729$ | $-0.0733$ |
| Channel 3-4 | $-2i$ | $-0.2343i$ | Same as Initial |
| Channel 3-5 | $\infty$ | $\infty$ | N/A (PMC) |
| Channel 3-6 | $\infty$ | $\infty$ | N/A (PMC) |
| Channel 4-5 | $\infty$ | $\infty$ | N/A (PMC) |
| Channel 4-6 | $\infty$ | $\infty$ | N/A (PMC) |

| Channel 5-6 | $2i$ | $0.2343i$ | Same as Initial |
|---|---|---|---|

### 4.2. Waveguide simulation results (six-port)

To validate the feasibility of the six-port ideal ZIM networks presented in Fig. 6 of Main Text, we demonstrate their waveguide-based implementations in this section. Figure S4 presents the simulated distributions of magnetic fields, confirming the successful replication of the sorting and synthesis functionalities observed in the ideal models. The geometric and electromagnetic parameters used in these simulations are provided in section 4.3.

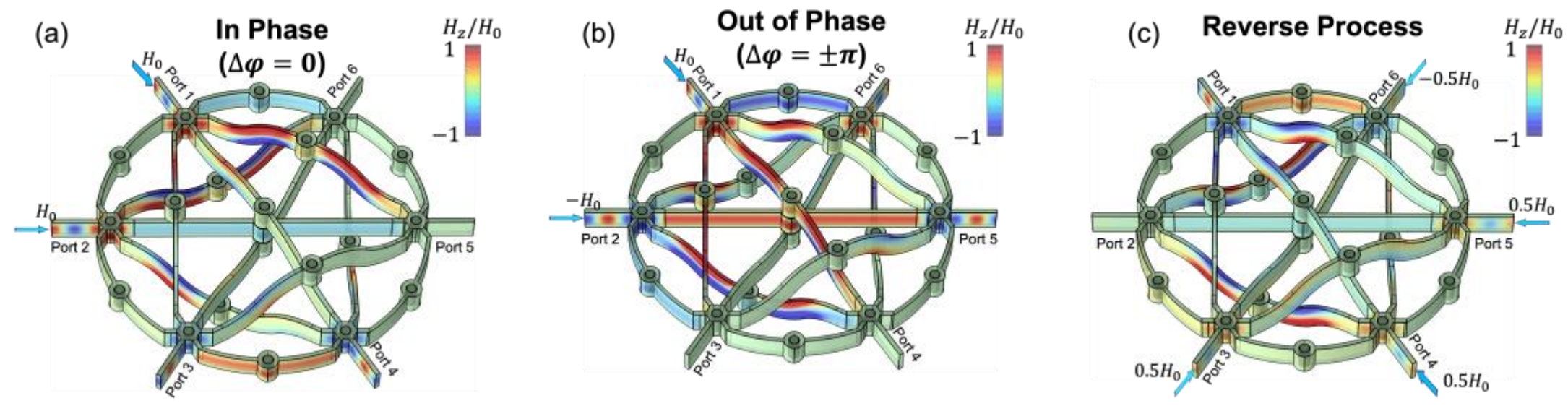

**Figure S4.** Full-wave simulation of the six-port waveguide-based mode sorter and synthesizer. [(a)-(c)] Simulated distributions of $H_z/H_0$ in an ideal ZIM network under different coherent excitations: (a) in-phase excitation of ports 1 and 2 (both with magnetic field $H_0$), (b) out-of-phase excitation of ports 1 and 2 ($H_0$ at port 1, $-H_0$ at port 2), and (c) simultaneous excitation of ports 3-6 with magnetic fields $0.5H_0$, $0.5H_0$, $0.5H_0$, and $-0.5H_0$, respectively.

### 4.3. Waveguide implementation parameters (six-port)

**Geometry:** The six-port waveguide model's topology is consistent with the ideal ZIM network model (section 4.1) and shares the same common parameters $h = 0.5\lambda_0$, $w = 0.1\lambda_0$ and $l_g = 0.25\lambda_0/\sqrt{1.1}$ as the four-port waveguide model (section 3.1). The dopant radius is $R_{d,mn} = 0.1\lambda_0$. The area $A_{mn}$ is defined as $A_{11} = A_{22} = A_{33} = A_{44} = A_{55} = A_{66} = 0.1697\lambda_0^2$, $A_{12} = A_{23} = A_{34} = A_{45} = A_{56} = A_{16} = 0.3405\lambda_0^2$, $A_{13} = A_{24} = A_{35} = A_{46} = A_{51} = A_{62} = 0.5563\lambda_0^2$, $A_{14} = A_{36} = 0.6714\lambda_0^2$, and $A_{25} = 0.6566\lambda_0^2$.

**Parameters:** The initial and refined dopant parameters for simulation models in Fig. S4 are provided in Table S8.

**Table S8: Dopant permittivities for the waveguide based six-port mode sorter**

| ZIM units | Target $\xi_{mn}$ | Initial $\varepsilon_{d,mn}$ | Refined $\varepsilon_{d,mn}$ |
|---|---|---|---|
| Node 1 | $-1$ | 17.5838 | 17.5788 |
| Node 2 | $-1$ | 17.5838 | 17.5788 |
| Node 3 | $-1+i$ | $17.5675+0.1767i$ | $17.5619+0.1749i$ |
| Node 4 | $-1+i$ | $17.5675+0.1767i$ | $17.5619+0.1749i$ |
| Node 5 | 0 | 17.7801 | 17.7743 |
| Node 6 | 0 | 17.7801 | 17.7743 |
| Channel 1-2 | $\infty$ | 15.6490 | 15.6490 |
| Channel 1-3 | $-2$ | 16.2112 | 16.1020 |
| Channel 1-4 | $-2$ | 16.1162 | 16.0901 |
| Channel 1-5 | $-2$ | 16.2112 | 16.1020 |
| Channel 1-6 | 2 | 16.7477 | 16.7410 |
| Channel 2-3 | $-2$ | 16.5574 | 16.5509 |
| Channel 2-4 | $-2$ | 16.2112 | 16.1020 |
| Channel 2-5 | 2 | 16.1745 | 16.1708 |
| Channel 2-6 | $-2$ | 16.2112 | 16.1020 |
| Channel 3-4 | $-2i$ | $16.6347-0.0935i$ | $16.6309-0.0940i$ |
| Channel 3-5 | $\infty$ | 15.6490 | 15.6490 |
| Channel 3-6 | $\infty$ | 15.6490 | 15.6490 |
| Channel 4-5 | $\infty$ | 15.6490 | 15.6490 |
| Channel 4-6 | $\infty$ | 15.6490 | 15.6490 |
| Channel 5-6 | $2i$ | $16.6347+0.0935i$ | $16.6309+0.0940i$ |

## 5. Planar implementation of the four-port coherent spatial mode demultiplexer

For small-scale routing networks (e.g., $N \leq 4$), the routing functionalities can indeed be realized using a strictly planar layout without waveguide crossings. To illustrate this, Figure S5 presents a planar implementation of the four-port coherent spatial mode demultiplexer corresponding to Figure 5 of the main text. Using the same target scattering matrix, and therefore the same characteristic parameter $\xi_{mn}$, as Figure 5, the planar layout is demonstrated with the ideal ZIM model. The structural parameters are $w = 0.5\lambda_0$, $l_g = 0.25\lambda_0$, $A_{11} = A_{22} = A_{33} = A_{44} = A_{12} = A_{13} = A_{14} = 0.25\lambda_0^2$, $A_{23} = A_{34} = \lambda_0^2$, $A_{24} = 2.5\lambda_0^2$, and $z_p = z_g = 1$. The relative permeability of each ZIM region is obtained from $\mu_{mn} = \xi_{mn} z_p w / k_0 A_{mn}$. The simulated field distributions confirm that the same routing functionality is preserved: in-phase excitation directs the output entirely to Port 3, whereas out-of-phase excitation redirects it entirely to Port 4.

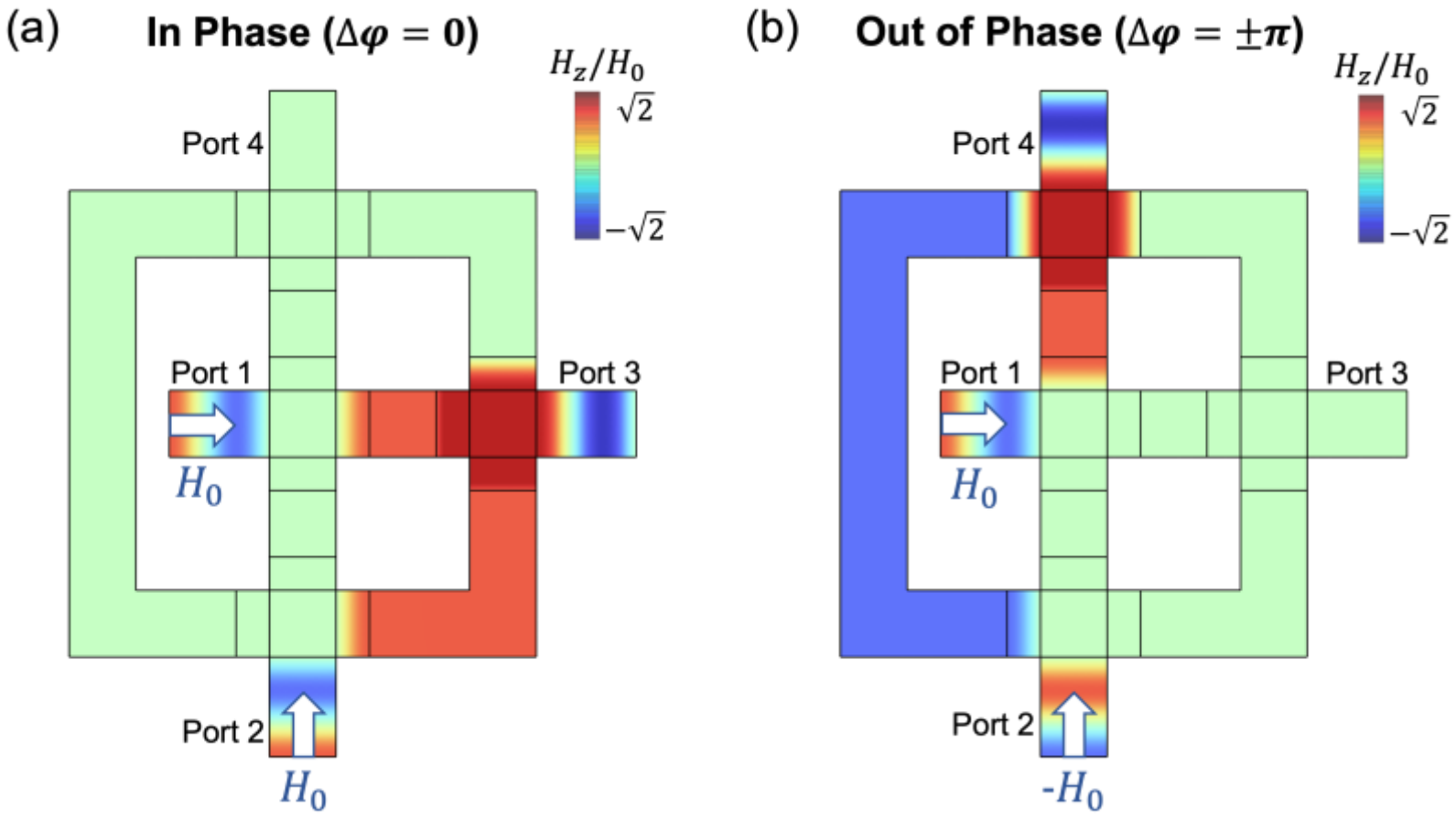

**Figure S5.** Simulated normalized magnetic-field distributions, $H_z/H_0$, in the coherent spatial mode demultiplexer implemented in a planar layout for (a) in-phase inputs ($\Delta\varphi = 0$) and (b) out-of-phase inputs ($\Delta\varphi = \pm\pi$).